\documentclass[journal,twoside]{IEEEtran}
\usepackage{amsmath}
\usepackage{amsthm}
\usepackage{amsfonts}
\usepackage{amssymb}
\usepackage{mathrsfs}
\usepackage{cases}
\usepackage{cite}
\usepackage{graphicx}
\usepackage{mathrsfs}
\usepackage{subfigure}
\usepackage{epstopdf}
\usepackage{color}
\usepackage{caption}

\usepackage{enumitem}
\usepackage{setspace}
\usepackage{bm}

\theoremstyle{remark}
\newtheorem{theorem}{\hspace{1em}Theorem}
\newtheorem{lemma}{\hspace{1em}Lemma}
\newtheorem{remark}{\hspace{1em}Remark}

\begin{document}

\title{Multiple Antennas Secure Transmission under Pilot Spoofing and Jamming Attack
\thanks{$^\dagger$ The authors are with the
School of Electronics and Information Engineering, and also with the Ministry
of Education Key Lab for Intelligent Networks and Network Security, Xi'an Jiaotong
University, Xi'an, 710049, Shaanxi, P. R. China. Email: {\tt
xjbswhm@gmail.com,  xjtu-huangkw@outlook.com}.
}
\thanks{$^\ddagger$ The author is with the School of Engineering, Nazarbayev University, Astana 010000, Kazakhstan. Email: {\tt theodoros.tsiftsis@nu.edu.kz}.}
\author{ Hui-Ming Wang$^\dagger$, \emph{Senior Member, IEEE}, \hspace{0.05in} Ke-Wen Huang$^\dagger$, \hspace{0.05in} Theodoros A. Tsiftsis$^\ddagger$, \emph{Senior Member, IEEE}
}
}
\maketitle

\begin{abstract}
    Transmitter-side channel state information (CSI) of the legitimate destination plays a critical role in physical layer secure transmissions.
    However, channel training procedure is vulnerable to the pilot spoofing attack (PSA) or pilot jamming attack (PJA) by an active eavesdropper (Eve), which inevitably results in severe private information leakage. 
    In this paper, we propose a random channel training (RCT)  based secure downlink transmission framework for a time division duplex (TDD) multiple antennas base station (BS). 
    In the proposed RCT scheme, multiple orthogonal pilot sequences (PSs) are simultaneously allocated to the legitimate user (LU), and the LU randomly selects one PS from the assigned PS set to transmit. 
    Under either the PSA or PJA, we provide the detailed steps for the BS to identify the PS transmitted by the LU, and to simultaneously estimate channels of the LU and Eve. 
    The probability that the BS makes an incorrect decision on the PS of the LU is analytically investigated.  Finally, closed-form secure beamforming (SB) vectors are designed and optimized to enhance the secrecy rates during the downlink transmissions. 
    Numerical results show that the secrecy performance is greatly improved compared to the conventional channel training scheme wherein only one PS is assigned to the LU.
\end{abstract}


\IEEEdisplaynontitleabstractindextext
\IEEEpeerreviewmaketitle
\section{Introduction}
Due to the openness of the wireless environments, wireless signals are vulnerable to be intercepted by a malicious eavesdropper (Eve). Since the pioneering work in \cite{A.D.Wyner1975}, physical layer security (PLS) approach has attracted increasing attention and has been regarded as an important complement to traditional cryptography
techniques to protect the secrecy of wireless transmissions \cite{Y.每W.P.Hong2013SPM,Y.Liu2017CST}.
Recently, by exploiting the extra spatial degrees of freedom provided by multiple antennas, multiple-input and multiple-output (MIMO) techniques have been applied to PLS to further enlarge  the secrecy capacity \cite{Y.每W.P.Hong2013SPM,X.Chen2017CST}.

Exploiting multiple antennas to facilitate PLS, secure beamforming (SB) and artificial noise (AN) aided transmission are two well-known approaches that have been extensively investigated in various contexts, e.g.,
in point to point MIMO systems \cite{T.Liu2009TIT,A.Khisti2010TITNov,
F.Oggier2011TIT},
in cooperative/relay systems
\cite{L.Dong2010TSP,H.-M.Wang2012TSP,H.每M.Wang2015CM,X.Wang2013TVT},
and in multi-cell and massive MIMO systems \cite{J.Zhu2014TWC,Wang2016Physical,HMWang2016Book,J.Zhu2016TWC}.
In all these works, transmitter-side channel state information (CSI) is of great importance in designing SB and AN signals. In practice, CSI is usually obtained by channel training, i.e., transmitting previously known pilot sequence (PS) for channel estimation. However, in most of existing works on the design of SB and AN signals, the transmitter-side CSI is directly assumed in perfect or imperfect forms, without considering the channel training phase.

In fact, intelligent active Eve can greatly improve its wiretapping performance by attacking the channel training phase of legitimate links, e.g., pilot spoofing attack (PSA) and pilot jamming attack (PJA). In PSA, Eve transmits the same PS as the legitimate user (LU) during the channel training phase. As a result, the channel estimation at the BS becomes a combination of the LU's and Eve's channels, and the designed beamforming vector is directed to both the LU and Eve, which results in serious private information leakage \cite{X.Zhou2012TWC}. In PJA, Eve transmits randomly generated interference to reduce the accuracy of the channel estimation.
As proved in \cite{Y.O.Basciftci2015CNS}, if Eve jams the channel training phase of a legitimate link, then the secure degree of freedom (SDOF) drops to zero. In viewing of the severe threat posed by these attacks, in this paper, we investigate how to overcome the PSA and PJA during the channel training phase and to achieve secure data transmissions via multiple antennas.

\subsection{Related works}
\subsubsection{Detection of PSA}
Many works have been focused on detecting the PSA
\cite{J.K.Tugnait,J.K.Tugnait2015WCL,
J.M.Kang2015VTC,Q.Xiong2015TIFS,D.Kapetanovic2013PIMRC}.
In \cite{D.Kapetanovic2013PIMRC}, a phase-shift keying based random training scheme was proposed to detect the PSA.
The authors of \cite{Q.Xiong2015TIFS} proposed to compare the signal power received at the base station (BS) and the LU to determine whether a PSA exists.
A generalized likelihood ratio based hypothesis testing method is derived in \cite{J.M.Kang2015VTC} to detect the PSA.
In \cite{J.K.Tugnait,J.K.Tugnait2015WCL}, the PS was transmitted along with randomly generated binary sequences, and the PSA can be detected by checking the \emph{rank} of the signal space.
{Though these works showed good performance on detecting the PSA, they did not provide a method to combat with it.}

\subsubsection{Secure transmission under PSA}
Secure transmission under the PSA was investigated in \cite{Q.Xiong2016TIFS,J.Xie2017ICC,Y.Wu2016TIT}.
In \cite{Q.Xiong2016TIFS}, a method utilizing the channel reciprocity of time division duplex (TDD) systems was developed to estimate the channels of the LU and Eve, based on which, SB was designed to protect data transmissions.
{The main drawback of this scheme  is that it requires an extra downlink training procedure which reduces the spectrum efficiency.}
In \cite{J.Xie2017ICC}, the authors proposed a two-stage channel training scheme to detect the PSA and estimate the legitimate and illegitimate channels for SB design. {However, the orthogonality of the PSs may be destroyed because different powers are used to transmit different parts of the PS, which means the scheme may only suit to single-user systems.}
In \cite{Y.Wu2016TIT}, the authors combated with the PSA by utilizing the different spatial channel statistics of the LU and Eve. {However, we observe that in \cite{Y.Wu2016TIT}, Eve can always transmit the same PS as the LU, which means that the estimation of the legitimate channel (the channel of the LU) is always contaminated by the illegitimate channel (the channel of Eve), and thus is of low accuracy.}

\subsubsection{Secure transmission under PJA}
Jamming attack and its countermeasures were investigated in \cite{T.T.Do2017WCL,H.Pirzadeh2016WCL,T.T.Do2018TIFS,H.AkhlaghpasandWCL}.
As shown in \cite{H.Pirzadeh2016WCL},
by properly allocating the jamming power during the channel training and data transmission phase, the spectral efficiency can be significantly degraded by a smart jammer.
In \cite{H.AkhlaghpasandWCL}, a GLLR based method was proposed to detect the jamming attack.
The authors of \cite{T.T.Do2017WCL} proposed to retransmit the PS to improve the accuracy of the channel estimation when the detected jamming power is large.
In \cite{T.T.Do2018TIFS}, the author proposed to estimate the jamming channel by exploiting a purposely unused PS, and  it was shown that with the estimated jamming channel, the average transmission rates were greatly increased.
Though these works have made great contributions on eliminating the effects of the jamming signals and enhancing the reliability of the received signals at the legitimate receiver, they did not consider the private information leakage at the malicious Eve due to the contaminated channel estimation.
To our knowledge, the research to ensure secure downlink transmission under PJA is still sparse. It was shown in \cite{Y.O.Basciftci2015CNS} that PJA during the channel training phase will drive the SDOF to zero. To combat with PJA, the authors  proposed  to  share the PS between the BS and the LU secretly through a cryptography key-based method, which violated the basic principle of the PLS.

\subsection{Motivations and Contributions}
Motivated by the severe threat posed by the PSA and PJA during the channel training phase,
in this paper,
we propose a framework for secure downlink transmissions in a multiple-antenna TDD system, incorporating uplink channel training, channel estimation and downlink secure beamforming design.
More specifically,
we propose a random channel training (RCT) scheme which is capable of combating with both the PSA and PJA.
Different from conventional channel training scheme wherein each user is assigned with only one PS, e.g., in \cite{J.Zhu2014TWC} and \cite{J.Zhu2016TWC},
the RCT scheme simultaneously assigns \emph{multiple} orthogonal PSs to the LU that is under attack. During the channel training phase, the LU \emph{randomly} chooses one PS from the assigned PS set to transmit.
Neither Eve nor the BS knows which PS will be transmitted by the LU.
After receiving the training signals, the BS  can distinguish the PS of the LU by exploiting the spatial channel statistics. Then simultaneous estimation of the legitimate and illegitimate channels is performed, and finally the SB vector is designed and optimized accordingly to maximize the average secrecy rate.
The main contributions of this paper can be summarized as follows:
\begin{itemize}[leftmargin=*]
    \item We propose a RCT scheme for the uplink channel training to combat with both the PSA and PJA. The RCT scheme is capable of simultaneously estimating the legitimate and illegitimate channels.
    We provide corresponding method for the BS to identify which PS is transmitted by the LU, without requiring any key-based information exchange. The error decision rate (EDR), defined as the probability that the BS makes an incorrect decision on the PS of the LU, has also been investigated, and tractable analytical expressions are derived for numerical evaluations.

    \item The SB vector is designed and optimized to improve the secrecy rate. Under the PSA, we obtain a tractable approximation of the average secrecy rate, which takes the channel estimation errors into consideration. Closed-form SB vector is derived which maximizes the approximate average secrecy rate. Under the PJA, we limit the signal leakage to Eve by only transmitting the signals on the null-space of the estimated illegitimate channel direction, and closed-form SB vector which maximizes the average SNR of the LU is also obtained.

   \item Analytical and numerical results show that the EDR of the proposed RCT scheme is very low.  The proposed framework greatly increases the secrecy rate during the downlink data transmission phase, compared with the existing schemes. Besides, the scheme can also be flexibly applied to multi-user downlink transmissions.

\end{itemize}
\subsection{Paper Organization and Notations}
The rest of the paper is organized as follows: In Section II, the proposed RCT scheme and the attacking schemes of Eve are introduced. In Section III--Section V, we introduce the uplink channel training, channel estimation and  downlink secure beamforming design under the PSA, respectively, in details. In Section IV, the counterpart to PJA is discussed. Numerical results are provided in Section VII. In Section VIII, we conclude the paper.

\emph{Notations:}$(\cdot)^H$ denotes the hermitian transpose. $\bm{I}_n$ denotes a $n$-by-$n$ identity matrix. $\left|\cdot\right|$ denotes the determinant, the absolute value or the cardinality. $\left\|\cdot\right\|$ denotes the $l_2$ norm.  $\mathcal{P}\left\{\cdot\right\}$ and $\mathbb{E}\left\{\cdot\right\}$ denote the probability and mathematical expectation.
$\mathbb{C}^{N\times M}$ denotes the space of $N$-by-$M$ complex-valued matrices. $\mathbb{CN}\left(\bm{\nu},\bm{\Sigma}\right)$ with $\bm{\nu}\in\mathbb{C}^{n\times 1}$ and $\bm{\Sigma}\in \mathbb{C}^{n\times n}$, $\mathbb{E}\left(\lambda\right)$, and $\mathbb{G}\left(\alpha,\beta\right)$ represent the complex-valued Gaussian distribution, the exponential distribution, and the Gamma distribution, respectively, with their probability density functions (PDFs) given by $\frac{1}{\pi^n|\bm{\Sigma}|}
\mathrm{e}^{-\left(\bm{x}-\bm{\nu}\right)^H\bm{\Sigma}^{-1}\left(\bm{x}-\bm{\nu}\right)}$, $\frac{\mathrm{e}^{x/\lambda}}{\lambda}$, and $\frac{x^{\alpha-1}\mathrm{e}^{x/\beta}}{\beta^\alpha\Gamma\left(\alpha\right)}$, respectively.
$\Gamma\left(M\right)$ and $\Gamma\left(M,x\right)$ are the Gamma function and the upper incomplete Gamma function \cite{I.S.Gradshteyn2007Table}, respectively. $\mathbb{I}_{\left\{\cdot\right\}}$ is the indicator function. $\bm{u}_{\mathrm{max}}\left\{\cdot\right\}$ denotes the normalized eigenvector of the largest eigenvalue of a matrix. $\left\langle\cdot,\cdot\right\rangle$ denotes the inner product. Block diagonal matrix is denoted by $\mathrm{diag}\left(\cdot,\cdots,\cdot\right)$.

\section{System model and random training scheme}
We consider the downlink secure transmission in a TDD system which consists of a multiple-antenna BS, a
single-antenna LU, and a single-antenna active Eve.
The whole transmission is divided into two phases, i.e., uplink channel training phase (UCTP) and downlink data transmission phase (DDTP).
During the UCTP, the LU transmits a PS to enable the BS to estimate the channel.
During the DDTP, based on the estimated channel, the BS forms a beam towards the LU to transmit confidential messages.
If Eve attacks the UCTP, e.g., PSA or PJA, the BS will obtain a misled channel estimation, and the downlink beam formed by the BS will not be aligned with the legitimate channel, which leads to serious information leakage. In the following, we introduce our RCT scheme to combat with the PSA and PJA.

\subsection{Random channel training scheme}
\begin{figure}[!t]
\begin{center}
\includegraphics[width=3.2 in]{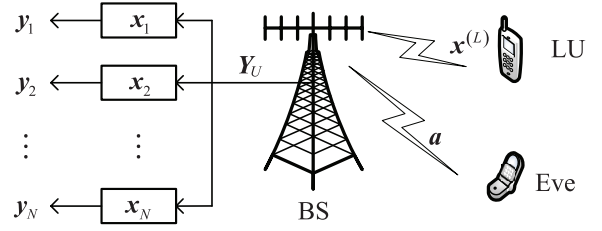}
\caption{Uplink Training Model.}\label{Fig:Model}
\end{center}
\vspace{-5mm}
\end{figure}

The procedure of the proposed RCT scheme is shown in Fig. \ref{Fig:Model}.
Different from conventional schemes where each user is only assigned with one PS, we simultaneously
allocate $N$ ($N>1$) orthogonal PSs to the LU \footnote{In this paper,
	we focus on a single-cell system with only one BS.
	In multi-cell environments, the PSs will be multiplexed in more than one cell, and the pilot contamination from other cells makes the problem mathematically much more complicated. We leave the multi-cell environments for future research.}.
Denote the set of PSs allocated to the LU as $\Phi\triangleq\{\bm{x}_n\}_{n=1}^N$, then we have
$\bm{x}_i^H\bm{x}_j=\tau\mathbb{I}_{\{i=j\}}$, where $\tau$ is the length of the PS.
During the UCTP, the LU \emph{randomly} chooses a PS from $\Phi$, denoted by $\bm{x}^{(L)}$, to perform the uplink channel training.
We emphasize that only the LU knows $\bm{x}^{(L)}$. Both Eve and the BS know the set $\Phi$ \footnote{$\Phi$ is assigned by the BS and shared between the BS and the LU, which is possible to be successfully intercepted by Eve. As the worst case, we assume that Eve perfectly knows $\Phi$.} but do not know which PS is $\bm{x}^{(L)}$.
For a given  instance of UCTP, the BS receives
\begin{align}
\bm{Y}_U \triangleq \sqrt{p_L}\bm{h}_L\left(\bm{x}^{(L)}\right)^H + \sqrt{p_E}\bm{h}_E\bm{a}^H + \bm{V},
\label{BSreceivedSignal}
\end{align}
where $p_L$ and $p_E$ are the powers of the LU and Eve, $\bm{h}_{L}\in\mathbb{C}^{M\times 1}\sim\mathbb{CN}\left(\bm{0},\bm{R}_L\right)$ and  $\bm{h}_{E}\in\mathbb{C}^{M\times 1}\sim\mathbb{CN}\left(\bm{0},\bm{R}_E\right)$ denote the instantaneous channels of the LU and Eve, $\bm{a}\in\mathbb{C}^{\tau\times 1}$ is the attacking signal sequence transmitted by Eve, which will be detailed later on, and $\bm{V}\in\mathbb{C}^{M\times\tau}$ is the noise with each element distributed as $\mathbb{CN}\left(0,\sigma_T^2\right)$.
Due to the fact that $\bm{R}_L$ and $\bm{R}_E$ change much slower than the instantaneous channels, we assume that the BS can obtain $\bm{R}_L$ and $\bm{R}_E$ in advance\footnote{Note that if $\mathrm{Tr}\left(\bm{R}_L\bm{R}_E\right)=0$, then the attack causes no impact on estimating LU's channel, and in this paper, we only consider the cases where $\mathrm{Tr}\left(\bm{R}_L\bm{R}_E\right)\gg0$, which means that the attacks cause serious impact on the channel estimation procedure.}.
Note that similar assumptions has also been adopted in \cite{Y.Wu2016TIT}.
After receiving $\bm{Y}_U$, the BS matches $\bm{Y}_U$ with the $N$ different orthogonal PSs in $\Phi$ as illustrated in Fig. 1, and obtains $N$ \emph{channel observations}, i.e.,
\begin{align}
\bm{y}_n & \triangleq \varphi\bm{Y}_U\bm{x}_n\nonumber \\
& = \mathbb{I}_{\left\{\bm{x}_n=\bm{x}^{(L)}\right\}}\bm{h}_L
+\varphi\sqrt{p_E}\bm{h}_E\bm{a}^H\bm{x}_n + \bm{z}_n,
\label{ObservationSetOriginal}
\end{align}
where $1\leq n\leq N$, $\varphi\triangleq\frac{1}{\tau\sqrt{p_L}}$, and $\bm{z}_n\triangleq \varphi \bm{V}\bm{x}_n \sim \mathbb{CN}\left(\bm{0},\sigma_z^2\bm{I}_M\right)$ with $\sigma_z^2\triangleq\frac{\sigma_T^2}{\tau p_L}$ is the equivalent Gaussian noise. Note that for $n\neq m$, $\bm{z}_n$ and $\bm{z}_m$ are mutually independent because $\bm{x}_n$ is orthogonal to $\bm{x}_m$.
Once obtaining the set of channel observations, i.e., $\Psi\triangleq\{\bm{y}_1,\bm{y}_2,\cdots,\bm{y}_N\}$, the BS needs to figure out which PS is $\bm{x}^{(L)}$ in order to estimate LU's channel, i.e, $\bm{h}_L$. Before presenting the details, we first discuss the attacking schemes of Eve in the following.

\subsection{Attacking Schemes of Eve}
In the RCT scheme described above, Eve is incapable of always transmitting the same PS as that transmitted by the LU due to the randomness.
We consider the following two possible attacking strategies of an active Eve,
\begin{itemize}
\item \emph{Pilot spoofing attack (PSA)}: Eve randomly selects several PSs from $\Phi$ and transmits a combination of them during the UCTP. Conventional PSA where Eve transmits the same PS as the LU \cite{X.Zhou2012TWC} can be viewed as a special case of the PSA defined here;
\item \emph{Pilot jamming attack (PJA)}: Eve transmits randomly generated Gaussian interference to degrade the accuracy of the channel estimation.
\end{itemize}
We discuss the results of these two kinds of attack in the following.
\subsubsection{PSA during UCTP}
Denote $\Phi_E\subseteq\Phi$ as the set of PSs selected by Eve during the UCTP.
Then, the attacking signal sequence transmitted by Eve is
\begin{align}
\label{PSAAttack}
\bm{a}_{\mathrm{PSA}} \triangleq \sum_{n=1}^N\mathbb{I}_{\left\{\bm{x}_n\in\Phi_E\right\}}\mathrm{e}^{\mathrm{j}\omega_n}\sqrt{\chi_n}\bm{x}_n,
\end{align}
where $\chi_n$, satisfying $\sum_{n=1}^N \mathbb{I}_{\left\{\bm{x}_n\in\Phi_E\right\}}\chi_n=1$, is the power factor for transmitting $\pmb{x}_n$, $\omega_n$ is a random phase shift when transmitting $\bm{x}_n$ which is unknown by the BS.
We will show that an extra random phase shift, i.e., $\mathrm{e}^{\mathrm{j}\omega_n}$, is necessary for Eve in Section \ref{Expample}.
In this paper, we assume that Eve uniformly allocate its power among different PSs, i.e., $\chi_n=\frac{1}{K}$ for $\forall \bm{x}_n\in\Phi_E$ where $K\triangleq\left|\Phi_E\right|$. This is because Eve does not have any prior knowledge on which PS will be transmitted by the LU, and hence each PS does not have any priority to gain more power than the others.
According to \eqref{PSAAttack}, $\bm{y}_n$ in \eqref{ObservationSetOriginal} becomes
\begin{align}
\bm{y}_n = \mathbb{I}_{\{\bm{x}_n=\bm{x}^{(L)}\}}\bm{h}_L
+\mathbb{I}_{\{\bm{x}_n\in\Phi_E\}}\mathrm{e}^{\mathrm{j}\omega_n}\beta_{\mathrm{PSA}}^{(K)}\bm{h}_E + \bm{z}_n,
\label{ObservationSetOriginalPSA}
\end{align}
where $\beta_{\mathrm{PSA}}^{(K)}\triangleq\sqrt{\frac{p_E}{Kp_L}}$.
For notational simplicity, define two non-overlapping subsets of $\Phi$ as
\begin{align}
\label{PSASub1}
\bar{\Phi}_{\mathrm{PSA}}^{(E)}
&\triangleq\left\{\bm{x}_n|\bm{x}_n\in\Phi_E,\bm{x}_n\neq\bm{x}^{(L)} \right\}\nonumber \\
&=\left\{\bm{x}_1^{(E)},\bm{x}_2^{(E)},\cdots,\bm{x}_{Q_E}^{(E)}\right\},\\
\label{PSASub2}
\bar{\Phi}_{\mathrm{PSA}}^{(F)}&\triangleq \Phi\setminus\left(\bar{\Phi}_{\mathrm{PSA}}^{(E)}\cup\{\bm{x}^{(L)}\}\right)\nonumber\\
&=
\left\{\bm{x}_1^{(F)},\bm{x}_2^{(F)},\cdots,\bm{x}_{Q_F}^{(F)}\right\},
\end{align}
where $Q_E\triangleq|\bar{\Phi}_{\mathrm{PSA}}^{(E)}|$ and $Q_F\triangleq|\bar{\Phi}_{\mathrm{PSA}}^{(F)}|$. Based on \eqref{PSASub1} and \eqref{PSASub2}, we can rewrite $\bm{y}_{n}$ in \eqref{ObservationSetOriginalPSA} as,
\begin{align}
\label{ObservationSetPSACase1}
\bm{y}_n&=\left\{
\begin{aligned}
\bm{y}^{(L)}, &\quad\quad\bm{x}_n=\bm{x}^{(L)},\\
\bm{y}_i^{(E)},&\quad\quad\bm{x}_n=\bm{x}_i^{(E)}\in \bar{\Phi}_{\mathrm{PSA}}^{(E)},\\
\bm{y}_i^{(F)}, &\quad\quad\bm{x}_n=\bm{x}_i^{(F)}\in \bar{\Phi}_{\mathrm{PSA}}^{(F)},
\end{aligned}\right. \\
\bm{y}^{(L)} &\triangleq \bm{h}_L + \mathbb{I}_{\{\bm{x}^{(L)}\in\Phi_E\}}\mathrm{e}^{\mathrm{j}\omega^{(L)}}\beta_{\mathrm{PSA}}^{(K)}\bm{h}_E + \bm{z}^{(L)}, \nonumber \\
\bm{y}_i^{(E)} &\triangleq \mathrm{e}^{\mathrm{j}\omega_i^{(E)}}\beta_{\mathrm{PSA}}^{(K)}\bm{h}_E + \bm{z}_i^{(E)},\nonumber\\
\bm{y}_i^{(F)} &\triangleq \bm{z}_i^{(F)},\nonumber
\end{align}
where
$\bm{z}^{(L)} \triangleq \sum_{n=1}^N\bm{z}_n\mathbb{I}_{\{\bm{x}_n=\bm{x}^{(L)}\}}$,
$\bm{z}_i^{(E)} \triangleq \sum_{n=1}^N\bm{z}_n\mathbb{I}_{\{\bm{x}_n=\bm{x}_i^{(E)}\}}$,
$\bm{z}_i^{(F)} \triangleq \sum_{n=1}^N\bm{z}_n\mathbb{I}_{\{\bm{x}_n=\bm{x}_i^{(F)}\}}$,
$\omega^{(L)} \triangleq \sum_{n=1}^N\omega_n\mathbb{I}_{\{\bm{x}_n=\bm{x}^{(L)}\}}$, and
$\omega_i^{(E)} \triangleq \sum_{n=1}^N\omega_n\mathbb{I}_{\{\bm{x}_n=\bm{x}_i^{(E)}\}}$.
According to \eqref{ObservationSetPSACase1}, we can divide $\Psi=\{\bm{y}_n\}_{n=1}^N$ into three non-overlapping subsets, i.e., $\Psi = \left\{\bm{y}^{(L)}\right\}\cup\Psi_{\mathrm{PSA}}^{(E)}\cup\Psi_{\mathrm{PSA}}^{(F)}$, where we have
$\Psi_{\mathrm{PSA}}^{(E)}
\triangleq\left\{\bm{y}_1^{(E)},\cdots,\bm{y}_{Q_E}^{(E)}\right\}$ and $\Psi_{\mathrm{PSA}}^{(F)}
\triangleq\left\{\bm{y}_1^{(F)},\cdots,\bm{y}_{Q_F}^{(F)}\right\}$.

We have to point out that the estimation of the legitimate channel is strongly affected by the realization of $\bm{x}^{(L)}$ and $\Phi_E$. Due to the fact that $\bm{x}^{(L)}$ and $\Phi_E$ are independently and randomly selected by the LU and Eve, respectively, we refer to the occurrence of a \emph{hit event} when $\bm{x}^{(L)}\in\Phi_E$, i.e., the PS transmitted by the LU is occasionally selected by Eve.

\subsubsection{PJA during UCTP}
During the UCTP, Eve transmits Gaussian random jamming signals
$\bm{a}_{\mathrm{PJA}}\sim \mathbb{CN}\left(\bm{0},\bm{I}_\tau\right)$.
Then, $\bm{y}_n$ in \eqref{ObservationSetOriginal} becomes
\begin{align}
\bm{y}_n = \varphi\bm{Y}_U\bm{x}_n = \mathbb{I}_{\left\{\bm{x}_n=\bm{x}^{(L)}\right\}}\bm{h}_L
+\mu_n\beta_{\mathrm{PJA}}\bm{h}_E + \bm{z}_n,
\label{ObservationSetOriginalJA}
\end{align}
where we have $\mu_n\triangleq\frac{1}{\tau}\bm{a}_{\mathrm{PJA}}^H\bm{x}_n^{(E)}
\sim\mathbb{CN}\left(0,\frac{1}{\tau}\right)$ and $\beta_{\mathrm{PJA}}\triangleq\sqrt{\frac{p_E}{p_L}}$.
Define $\bar{\Phi}_{\mathrm{PJA}}^{(E)}$ as a subset of $\Phi$ which is written as
\begin{align}
\bar{\Phi}_{\mathrm{PJA}}^{(E)}\triangleq\Phi\setminus\{\bm{x}^{(L)}\}
=\{\check{\bm{x}}_1^{(E)},\cdots,\check{\bm{x}}_{N-1}^{(E)}\}.
\end{align}
Similar to \eqref{ObservationSetPSACase1}, we rewrite the channel observations in \eqref{ObservationSetOriginalJA} under PJA  as
\begin{align}
\label{JAObservations}
\bm{y}_n &= \left\{
\begin{aligned}
\bm{y}^{(L)},&\quad\quad\bm{x}_n=\bm{x}^{(L)},\\
\bm{y}_i^{(E)},&\quad\quad \bm{x}_n=\check{\bm{x}}_i^{(E)}\in\bar{\Phi}_{\mathrm{PJA}}^{(E)},
\end{aligned}
\right.\\
\bm{y}^{(L)} &\triangleq\bm{h}_L+\mu^{(L)}\beta_{\mathrm{PJA}}\bm{h}_E + \bm{z}^{(L)},\nonumber\\
\bm{y}_i^{(E)} &\triangleq\mu_i^{(E)}\beta_{\mathrm{PJA}}\bm{h}_E + \bm{z}_i^{(E)},\nonumber
\end{align}
where we have
$\bm{z}_i^{(E)} \triangleq \sum_{n=1}^N\bm{z}_n\mathbb{I}_{\{\bm{x}_n=\check{\bm{x}}_i^{(E)}\}}$,
$\mu^{(L)} \triangleq \sum_{n=1}^N\mu_n\mathbb{I}_{\{\bm{x}_n=\bm{x}^{(L)}\}}$, and
$\mu_i^{(E)} \triangleq \sum_{n=1}^N\mu_n\mathbb{I}_{\{\bm{x}_n=\check{\bm{x}}_i^{(E)}\}}$.
According to \eqref{JAObservations}, we divide $\Psi$ under PJA  into two non-overlapping subsets, i.e., $\Psi=\left\{\bm{y}^{(L)}\right\}\cup\Psi_{\mathrm{PJA}}^{(E)}$, where we define
$\Psi_{\mathrm{PJA}}^{(E)}\triangleq
\left\{\bm{y}_1^{(E)},\cdots,\bm{y}_{N-1}^{(E)}\right\}$.

As we can see from \eqref{ObservationSetPSACase1}  and \eqref{JAObservations}, the BS obtains multiple channel observations, which contains both LU's and Eve's CSI.
A critical step for the BS to achieve secrecy transmission is to
figure out which channel observation is $\bm{y}^{(L)}$ so as to estimate the legitimate and illegitimate channels simultaneously.
We will discuss the detailed steps for the BS to identify $\bm{y}^{(L)}$, to estimate the channels, and to design the SB vector under the PSA, in Section \ref{PSAChannelEstimation}, Section \ref{ChannelEsPSA}, and Section \ref{PSASecureTransmission}, respectively.
Secure transmission under the PJA will be discussed in Section \ref{STJA} in a similar manner.
\begin{remark}
\label{Extension}
The proposed scheme can be easily extended to a multi-user system because we exploit the orthogonality among different PSs. More specifically, in a system with $U$ users, the set of PSs $\Phi$ can be divided into $U$ non-overlapping subsets, i.e., $\Phi^{(1)},\Phi^{(2)},\cdots,\Phi^{(U)}\subset\Phi$, with $\Phi^{(u)}$ allocated to the $u^{\mathrm{th}}$ user, and all the following discussions hold as well. In following part of this paper, unless specified, we always assume that there is only one user and all the PSs are allocated to the user, i.e., $N=\tau$.
\end{remark}

\section{Determining the PS of LU under PSA}
\label{PSAChannelEstimation}
In this section, we present the method for the BS to determine which PS is $\bm{x}^{(L)}$, or equivalently speaking, which channel observation is $\bm{y}^{(L)}$, under the PSA.
We also evaluate the detection performance in terms of the EDR, i.e., the probability that the BS makes an incorrect decision on $\bm{y}^{(L)}$.

We have to point out that, as shown in \eqref{ObservationSetPSACase1}, channel observations in $\Psi_{\mathrm{PSA}}^{(F)}$  contains neither LU's nor Eve's CSI, and therefore, as a pre-processing step, the BS needs to distinguish $\Psi_{\mathrm{PSA}}^{(C)}\triangleq \left\{\bm{y}^{(L)}\right\}\cup\Psi_{\mathrm{PSA}}^{(E)}$ from $\Psi_{\mathrm{PSA}}^{(F)}$.
To realize this, the BS can simply classify $\bm{y}_n$, ($1\leq n\leq N$), into $\Psi_{\mathrm{PSA}}^{(C)}$ or $\Psi_{\mathrm{PSA}}^{(F)}$ by comparing their powers with a predesigned threshold $\Lambda_C$, i.e.,$\left\|\bm{y}_n\right\|^2 \gtreqless_{\Psi_{\mathrm{PSA}}^{(F)}}^{\Psi_{\mathrm{PSA}}^{(C)}} \Lambda_C$.
We observe that the powers of $\bm{y}^{(L)}$ and $\bm{y}_{n}^{(E)}$ are much larger than $\bm{y}_{n}^{(F)}$ due to the fact that the noise floor of $\bm{z}_{n}^{(F)}$ is generally small, especially when $p_L$ and $\tau$ are large. Therefore, we make a reasonable assumption that the BS has already obtained the effective channel observation set $\Psi_{\mathrm{PSA}}^{(C)}$ correctly. For notational simplicity, we define $Q_C$ as the cardinality of $\Psi_{\mathrm{PSA}}^{(C)}$, i.e., $Q_C\triangleq|\Psi_{\mathrm{PSA}}^{(C)}|$.

Now, we provide our method for the BS to determine which channel observation in $\Psi_{\mathrm{PSA}}^{(C)}$ is $\bm{y}^{(L)}$.
We note that, the detection is highly related to the parameter $K$, which is, however, chosen by Eve and unknown by the BS.
For simplicity, In the  first step we discuss the cases where $K$ is known by BS, and classify our discussions according to the value of $K$ and whether or not a hit event occurs. After that, we extend our method to handle the cases when $K$ is unknown.

\subsection{$K=1$ and Eve successfully hits LU's PS}
In this cases, undoubtedly, we have $\Psi_{\mathrm{PSA}}^{(C)}= \left\{\bm{y}^{(L)}\right\}$ where $\bm{y}^{(L)} = \bm{h}_L+\mathrm{e}^{\mathrm{j}\omega^{(L)}}\beta_{\mathrm{PSA}}^{(1)}\bm{h}_E+\bm{z}^{(L)}$.
We note that this is the case of the conventional PSA and has been discussed in \cite{J.K.Tugnait2015WCL,J.M.Kang2015VTC,Q.Xiong2015TIFS,Q.Xiong2016TIFS,J.Xie2017ICC,Y.Wu2016TIT,J.K.Tugnait}.
\subsection{$K=1$ and Eve fails to hit LU's PS}
\label{K1FailHit}
In this case, we have $Q_C=2$. Without loss of generality, denote $\Psi_{\mathrm{PSA}}^{(C)}=\left\{\bm{y}_1, \bm{y}_2\right\}$.
To determine which one of $\bm{y}_1$ and $\bm{y}_2$ is $\bm{y}^{(L)}$, we formulate a hypothesis test problem, i.e.,
\begin{align}
\label{L1Psi2y1y2H0H1}
\begin{bmatrix}
\bm{y}_1 \\
\bm{y}_2 \\
\end{bmatrix}=
\left\{
\begin{aligned}
\begin{bmatrix}
\bm{y}^{(L)} \\
\bm{y}^{(E)} \\
\end{bmatrix}
=
\begin{bmatrix}
\bm{h}_L + \bm{z}^{(L)} \\
\mathrm{e}^{\mathrm{j}\omega^{(E)}}\beta_{\mathrm{PSA}}^{(1)}\bm{h}_E + \bm{z}^{(E)} \\
\end{bmatrix}, \mathcal{H}_0^{(1,2)},\\
\begin{bmatrix}
\bm{y}^{(E)} \\
\bm{y}^{(L)} \\
\end{bmatrix}
=
\begin{bmatrix}
 \mathrm{e}^{\mathrm{j}\omega^{(E)}}\beta_{\mathrm{PSA}}^{(1)}\bm{h}_E + \bm{z}^{(E)}\\
 \bm{h}_L + \bm{z}^{(L)}\\
\end{bmatrix}, \mathcal{H}_1^{(1,2)},
\end{aligned}\right.
\end{align}
where $\mathcal{H}_0^{(1,2)}$ represents that $\bm{y}_1=\bm{y}^{(L)}$ and $\mathcal{H}_1^{(1,2)}$ represents that $\bm{y}_2=\bm{y}^{(L)}$.
To determine $\bm{y}^{(L)}$, the logarithmic likelihood ratio (LLR) test is written as
\begin{align}
\label{L1PsiC2LikehoodRadioOriginal}
T_1^{(2)} \triangleq \ln p\left( \bm{g} |\mathcal{H}_{0}^{(1,2)} \right)-\ln p\left(\bm{g}|\mathcal{H}_{1}^{(1,2)} \right)
{\gtreqless}_{\mathcal{H}_1^{(1,2)}}^{\mathcal{H}_0^{(1,2)}} 0,
\end{align}
where $\bm{g}\triangleq\left[\bm{y}_1^H,\bm{y}_2^H\right]^H$, $p\left(\bm{g} |\mathcal{H}_{0}^{(1,2)}\right)$ and $p\left(\bm{g} |\mathcal{H}_{1}^{(1,2)}\right)$ are the PDFs of $\bm{g}$ conditioned on $\mathcal{H}_{0}^{(1,2)}$ and $\mathcal{H}_{1}^{(1,2)}$, respectively.
Note that if $\mathcal{H}_0^{(1,2)}$ is true, $\bm{y}_1$ and $\bm{y}_2$ are independently distributed as $\mathbb{CN}\left(\bm{0},\bm{R}_{L,z}\right)$ and
$\mathbb{CN}\left(\bm{0},\bm{R}_{E,z}^{(1)}\right)$, respectively, and if $\mathcal{H}_1^{(1,2)}$ is true, $\bm{y}_1$ and $\bm{y}_2$ are independently distributed as $\mathbb{CN}\left(\bm{0},\bm{R}_{E,z}^{(1)}\right)$ and
$\mathbb{CN}\left(\bm{0},\bm{R}_{L,z}\right)$, respectively, where we have $\bm{R}_{L,z}\triangleq\bm{R}_L+\sigma_z^2\bm{I}$ and
$\bm{R}_{E,z}^{(1)}\triangleq|\beta_{\mathrm{PSA}}^{(1)}|^2\bm{R}_E+\sigma_z^2\bm{I}$.
Therefore, \eqref{L1PsiC2LikehoodRadioOriginal} can be further simplified as
\begin{align}
T_1^{(2)} &= \bm{y}_1^H\left(\left(\bm{R}_{E,z}^{(1)}\right)^{-1} - \bm{R}_{L,z}^{-1}\right)\bm{y}_1
\nonumber \\
&\quad + \bm{y}_2^H\left(\bm{R}_{L,z}^{-1} - \left(\bm{R}_{E,z}^{(1)}\right)^{-1}\right)\bm{y}_2
{\gtreqless}_{\mathcal{H}_1^{(1,2)}}^{\mathcal{H}_0^{(1,2)}} 0.
\label{L1PsiC2LikehoodRadioFinal}
\end{align}
According to \eqref{L1PsiC2LikehoodRadioFinal}, the EDR in this case is given by
\begin{align}
\mathcal{P}_{\mathrm{EDR}}^{(1,2)}&\triangleq \mathcal{P}\left\{T_1^{(2)} < 0 | \mathcal{H}_0^{(1,2)} \right\} \nonumber \\
&= \mathcal{P}\left\{T_1^{(2)} >0 | \mathcal{H}_1^{(1,2)} \right\} \label{EDR12Original}.
\end{align}
The calculation of $\mathcal{P}_{\mathrm{EDR}}^{(1,2)}$ is provided in the following theorem.
\begin{theorem}
\label{Theorem:T12}
$\mathcal{P}_{\mathrm{EDR}}^{(1,2)}$ can be calculated as
\begin{align}
\label{PEDRT12}
\mathcal{P}_{\mathrm{EDR}}^{(1,2)}&
=\mathcal{P}\left\{ \bm{w}^H\bm{\Xi}\bm{w} < 0 \right\},
\end{align}
where $\bm{w}$ is a Gaussian random vector distributed as $\mathbb{CN}\left(\bm{0},\bm{I}_{2M}\right)$, and
\begin{align}
\label{DefinitionTheta}
\bm{\Xi}&\triangleq
\begin{bmatrix}
\bm{\Xi}_1-\bm{I}_M,&\bm{0}_{M\times M}\\
\bm{0}_{M\times M},&\bm{\Xi}_2-\bm{I}_M \\
\end{bmatrix},
\end{align}
with 
$\bm{\Xi}_1\triangleq\bm{R}_{L,z}^{\frac{1}{2}}
\left(\bm{R}_{E,z}^{(1)}\right)^{-1}
\bm{R}_{L,z}^{\frac{1}{2}}$ 
and 
$\bm{\Xi}_2\triangleq\left(\bm{R}_{E,z}^{(1)}\right)^{\frac{1}{2}}
\bm{R}_{L,z}^{-1}
\left(\bm{R}_{E,z}^{(1)}\right)^{\frac{1}{2}}$.
\end{theorem}
\begin{IEEEproof}
The detailed derivation of \eqref{DefinitionTheta} is provided in \eqref{ProofTheoremX} at the top of the next page,
\begin{figure*}[t]
\begin{align}
\mathcal{P}_{\mathrm{EDR}}^{(1,2)} &=\mathcal{P}\left\{\bm{y}_1^H\left(\left(\bm{R}_{E,z}^{(1)}\right)^{-1} - \bm{R}_{L,z}^{-1}\right)\bm{y}_1+ \bm{y}_2^H\left(\bm{R}_{L,z}^{-1} - \left(\bm{R}_{E,z}^{(1)}\right)^{-1}\right)\bm{y}_2<0|\mathcal{H}_0^{(1,2)}\right\}\nonumber \\
&= \mathcal{P}\left\{\bm{y}_1^H\bm{R}_{L,z}^{-\frac{1}{2}}\left(
\bm{\Xi}_1 - \bm{I}_M\right)\bm{R}_{L,z}^{-\frac{1}{2}}\bm{y}_1+ \bm{y}_2^H\left(\bm{R}_{E,z}^{(1)}\right)^{-\frac{1}{2}}\left(\bm{\Xi}_{2}- \bm{I}_M\right)\left(\bm{R}_{E,z}^{(1)}\right)^{-\frac{1}{2}}\bm{y}_2<0|\mathcal{H}_0^{(1,2)}\right\}\nonumber \\
&= \mathcal{P}\left\{\bm{w}_1^H\left(
\bm{\Xi}_1 - \bm{I}_M\right)\bm{w}_1+ \bm{w}_2\left(\bm{\Xi}_{2}- \bm{I}_M\right)\bm{w}_2^H<0\right\}=\mathcal{P}\left\{\bm{w}^H\bm{\Xi}\bm{w}<0\right\}.
\label{ProofTheoremX}
\end{align}
\hrulefill
\end{figure*}
where we have $\bm{w}_1\triangleq\bm{R}_{L,z}^{-\frac{1}{2}}\bm{y}_1$, $\bm{w}_2\triangleq\left(\bm{R}_{E,z}^{(1)}\right)^{-\frac{1}{2}}\bm{y}_2$,
$\bm{w}\triangleq\left[\bm{w}_1^H,\bm{w}_2^H\right]^H$, and $\bm{\Xi}_1$, $\bm{\Xi}_2$ and $\bm{\Xi}$ are defined in \eqref{DefinitionTheta}.
Note that conditioned on $\mathcal{H}_0^{(1,2)}$, $\bm{y}_1$ and $\bm{y}_2$ are independently distributed as $\mathbb{CN}\left(\bm{0},\bm{R}_{L,z}\right)$ and $\mathbb{CN}\left(\bm{0},\bm{R}_{E,z}^{(1)}\right)$, respectively. Therefore, we have $\bm{w}\sim\mathbb{CN}\left(\bm{0},\bm{I}_{2M}\right)$.
In fact, \eqref{PEDRT12} is the CDF of an indefinite quadratic form of a Gaussian random vector, the calculation of which is provided in Appendix \ref{QuadraticGaussian}.
\end{IEEEproof}
\subsection{$K=2$ and Eve successfully hits LU's PS}
\label{Expample}
In this case, we also have $Q_C=2$. Without loss of generality, denoting $\Psi_{\mathrm{PSA}}^{(C)}=\left\{\bm{y}_1,\bm{y}_2\right\}$, we have
\begin{align}
\label{L2Psi2y1y2H0}
\begin{bmatrix}
\bm{y}_1 \\
\bm{y}_2 \\
\end{bmatrix} =
\left\{
\begin{aligned}
\begin{bmatrix}
\bm{y}^{(L)} \\
\bm{y}^{(E)} \\
\end{bmatrix}=
\begin{bmatrix}
\bm{h}_L + \mathrm{e}^{\mathrm{j}\omega^{(L)}}\beta_{\mathrm{PSA}}^{(2)}\bm{h}_E + \bm{z}^{(L)} \\
\mathrm{e}^{\mathrm{j}\omega^{(E)}}\beta_{\mathrm{PSA}}^{(2)}\bm{h}_E + \bm{z}^{(E)} \\
\end{bmatrix},\mathcal{H}_0^{(2,2)}, \\
\begin{bmatrix}
\bm{y}^{(E)} \\
\bm{y}^{(L)} \\
\end{bmatrix}=
\begin{bmatrix}
\mathrm{e}^{\mathrm{j}\omega^{(E)}}\beta_{\mathrm{PSA}}^{(2)}\bm{h}_E + \bm{z}^{(E)}\\
\bm{h}_L + \mathrm{e}^{\mathrm{j}\omega^{(L)}}\beta_{\mathrm{PSA}}^{(2)}\bm{h}_E + \bm{z}^{(L)}\\
\end{bmatrix},\mathcal{H}_1^{(2,2)},
\end{aligned}\right.
\end{align}
where $\mathcal{H}_0^{(2,2)}$ represents that $\bm{y}_1=\bm{y}^{(L)}$, and $\mathcal{H}_1^{(2,2)}$ represents the opposite.
Before determining which one of $\mathcal{H}_1^{(2,2)}$ and $\mathcal{H}_2^{(2,2)}$ is true, we have the following remark.
\begin{remark}
In fact, \eqref{L2Psi2y1y2H0} provides us an explicit explanation on why random phase shifts $\mathrm{e}^{\mathrm{j}\omega_n}$ are necessary when Eve transmits multiple PSs. If the random phase shifts are absent, then once the BS knows which channel observation is $\bm{y}^{(L)}$, it can cancel the impact of $\bm{h}_E$ on $\bm{y}^{(L)}$ by subtracting $\bm{y}^{(E)}$ from $\bm{y}^{(L)}$ and obtain an uncontaminated version of channel observation of $\bm{h}_L$, i.e., $\check{\bm{y}}^{(L)}\triangleq\bm{y}^{(L)}-\bm{y}^{(E)}=\bm{h}_L + \bm{z}^{(L)}-\bm{z}^{(E)}$. Estimating $\bm{h}_L$ from $\check{\bm{y}}^{(L)}$ provides more accurate result than from $\bm{y}^{(L)}$, which is generally not expected by the Eve.
\end{remark}

Conditioned on $\mathcal{H}_{0}^{(2,2)}$, we have $\bm{g}\sim\mathbb{CN}\left(\bm{0},\bm{R}_{g,0}\left(\omega\right)\right)$, and conditioned on $\mathcal{H}_{1}^{(2,2)}$, we have $\bm{g}\sim\mathbb{CN}\left(\bm{0},\bm{R}_{g,1}\left(\omega\right)\right)$, where $\omega\triangleq\omega^{(L)}-\omega^{(E)}$, and
\begin{subequations}
\label{K2gDistribution}
\begin{align}
\bm{R}_{g,0}\left(\omega\right)&\triangleq
\begin{bmatrix}
\bm{R}_{L,E,z}^{(2)} & \mathrm{e}^{\mathrm{j}\omega}\bm{R}_{E}^{(2)} \\
\mathrm{e}^{-\mathrm{j}\omega}\bm{R}_{E}^{(2)} & \bm{R}_{E,z}^{(2)} \\
\end{bmatrix},\\
\bm{R}_{g,1}\left(\omega\right)&\triangleq
\begin{bmatrix}
\bm{R}_{E,z}^{(2)} & \mathrm{e}^{-\mathrm{j}\omega}\bm{R}_{E}^{(2)} \\
\mathrm{e}^{\mathrm{j}\omega}\bm{R}_{E}^{(2)} & \bm{R}_{L,E,z}^{(2)} \\
\end{bmatrix},
\end{align}
\end{subequations}
with $\bm{R}_{E}^{(2)}\triangleq|\beta_{\mathrm{PSA}}^{\left(2\right)}|^2\bm{R}_E$ and
$\bm{R}_{L,E,z}^{(2)}\triangleq\bm{R}_{L,z}+\bm{R}_{E}^{(2)}$.
Now, we provide our method to distinguish $\bm{y}^{(L)}$ from $\bm{y}^{(E)}$. Since the value of $\omega^{(L)}$ and $\omega^{(E)}$ are unknowns, using LLR test to determine $\bm{y}^{(L)}$ is not applicable. In the following, we provide two methods to determine $\bm{y}^{(L)}$.
\subsubsection{Power comparison based method}
We observe that $\bm{h}_L$ and $\bm{h}_E$ are mutually independent random vectors, and thus,
the power of $\bm{y}^{(L)}$ is expected to be larger than $\bm{y}^{(E)}$.
Based on this observation, we propose to simply compare the powers of $\bm{y}_1$ and $\bm{y}_2$, i.e.,
\begin{align}
\tilde{T}_2^{(2)}\triangleq\left\|\bm{y}_1\right\|^2-\left\|\bm{y}_2\right\|^2 {\gtreqless}_{\mathcal{H}_1^{(2,2)}}^{\mathcal{H}_0^{(2,2)}} 0.
\label{L2PsiC2LikehoodRadio}
\end{align}
According to \eqref{L2PsiC2LikehoodRadio}, the EDR is given by
\begin{align}
\tilde{\mathcal{P}}_{\mathrm{EDR}}^{(2,2)}&\triangleq
\mathcal{P}\left\{\tilde{T}_2^{(2)}>0|\mathcal{H}_1^{(2,2)}\right\}
=\mathcal{P}\left\{\tilde{T}_2^{(2)}<0|\mathcal{H}_0^{(2,2)}\right\}\nonumber \\
&=\mathcal{P}\Bigg\{\left\|\bm{h}_L + \mathrm{e}^{\mathrm{j}\omega^{(L)}}\beta_{\mathrm{PSA}}^{(2)}\bm{h}_E + \bm{z}^{(L)}\right\|^2
\nonumber\\
&\quad\quad\quad-\left\|\mathrm{e}^{\mathrm{j}\omega^{(E)}}\beta_{\mathrm{PSA}}^{(2)}\bm{h}_E + \bm{z}^{(E)}\right\|^2<0\Bigg\}.
\label{L2PsiC2ErrorRate}
\end{align}
We note that the calculation of \eqref{L2PsiC2ErrorRate} is difficult for two reasons:
(1) the joint PDF of $\left(\bm{y}_1,\bm{y}_2\right)$ is unknown due to the unknown parameters $\omega^{(L)}$ and $\omega^{(E)}$, and (2) though the marginal distribution of $\bm{y}_1$ and $\bm{y}_2$ are known, they are correlated with each other. Fortunately, we successfully obtain the following theorem to provide a tractable expression for
numerical evaluation of \eqref{L2PsiC2ErrorRate}.
\begin{theorem}
\label{Theorem:T22}
The EDR in \eqref{L2PsiC2ErrorRate} is irrelevant to $\omega^{(L)}$ and $\omega^{(E)}$, and can be written as
\begin{align}
\tilde{\mathcal{P}}_{\mathrm{EDR}}^{(2,2)}
=\mathcal{P}\left\{\bm{\alpha}^H\bm{W}\bm{\alpha}<0\right\}
\label{S6}
\end{align}
where $\bm{\alpha}\sim\mathbb{CN}\left(\bm{0},\bm{I}_{2M}\right)$ and $\bm{W}$ is an indefinite Hermitian matrix defined in Appendix \ref{Appendix:ProofT22}.
\end{theorem}
\begin{IEEEproof}
Please refer to Appendix \ref{Appendix:ProofT22}. Besides, we provide a mathematical expression for numerical evaluation of
\eqref{S6} in Appendix \ref{QuadraticGaussian}.
\end{IEEEproof}

\subsubsection{Generalized logarithmic likelihood ratio (GLLR) based method}
GLLR is an extension of regular LLR, which is usually used to handle hypothesis test problems with unknown parameters \cite{H.L.VanTrees1968}. The GLLR test for \eqref{L2Psi2y1y2H0} is given by
\begin{align}
\label{DecisionCriteronT22}
T_2^{(2)} &\triangleq
\ln \left(\max_{\omega^{(L)},\omega^{(E)}} p\left(\bm{g} |\mathcal{H}_{0}^{(2,2)} \right)\right)
\nonumber \\
&\quad-
\ln \left(\max_{\omega^{(L)},\omega^{(E)}} p\left(\bm{g} |\mathcal{H}_{1}^{(2,2)} \right)\right)
{\gtreqless}_{\mathcal{H}_1^{(2,2)}}^{\mathcal{H}_0^{(2,2)}} 0,
\end{align}
where for $i\in\{0,1\}$, $p\left(\bm{g} |\mathcal{H}_{i}^{(2,2)}\right)$ is the PDF of $\bm{g}$ conditioned on $\mathcal{H}_{i}^{(2,2)}$.
Based on  \eqref{K2gDistribution},  $T_2^{(2)}$ in \eqref{DecisionCriteronT22} can be further simplified as
\begin{align}
\label{GLLRTEST}
T_2^{(2)} &= \left(\min_{\omega}~\bm{y}_1^H\bm{B}\bm{y}_1+\bm{y}_2^H\bm{A}\bm{y}_2
+ 2\mathrm{Re}\left\{\mathrm{e}^{-\mathrm{j}\omega}\bm{y}_1^H\bm{C}^H\bm{y}_2\right\}\right)\nonumber\\
&\quad -
\left(\min_{\omega}~\bm{y}_1^H\bm{A}\bm{y}_1+\bm{y}_2^H\bm{B}\bm{y}_2
+ 2\mathrm{Re}\left\{\mathrm{e}^{\mathrm{j}\omega}\bm{y}_1^H\bm{C}\bm{y}_2\right\}\right)\nonumber\\
&=\bm{y}_1^{H}\left(\bm{B} - \bm{A}\right)\bm{y}_1 + \bm{y}_2^{H}\left(\bm{A} - \bm{B}\right)\bm{y}_2
\nonumber \\
&\quad+ 2\left|\bm{y}_1^{H}\bm{C}\bm{y}_2\right| - 2\left|\bm{y}_1^{H}\bm{C}^H\bm{y}_2\right|,
\end{align}
where $\bm{A}$, $\bm{B}$, and $\bm{C}$ are three $M$-dimensional square matrices satisfying
\begin{align}
\begin{bmatrix}
\bm{A} & \bm{C} \\
\bm{C}^H & \bm{B} \\
\end{bmatrix} =
\begin{bmatrix}
\bm{R}_{L,E,z}^{(2)} & \bm{R}_{E}^{(2)} \\
\bm{R}_{E}^{(2)} & \bm{R}_{E,z}^{(2)} \\
\end{bmatrix}^{-1}.
\end{align}
According to \eqref{DecisionCriteronT22}, the EDR is given by
\begin{align}
\mathcal{P}_{\mathrm{EDR}}^{(2,2)}
\triangleq\mathcal{P}\left\{T_2^{(2)}<0|\mathcal{H}_{0}^{(2,2)}\right\}=\mathcal{P}\left\{T_2^{(2)}>0|\mathcal{H}_{1}^{(2,2)}\right\}.
\nonumber
\end{align}
In the simulation part, we will show that $\mathcal{P}_{\mathrm{EDR}}^{(2,2)}$ is generally small, which means that the BS can successfully recognize $\bm{y}^{(L)}$ with high probability.
Note that the calculation of $T_2^{(2)}$ requires an inverse operation with respect to a $2M-dimensional$ matrix, therefore, the GLLR based method is computationally more complex than the power comparison based method.

\subsection{$K=2$ but Eve fails to hit LU's PS, or $K>2$}
\label{Section:ConterAttack:LargeL}
\begin{figure}[t]
	\centering
	\includegraphics[width=3.3 in]{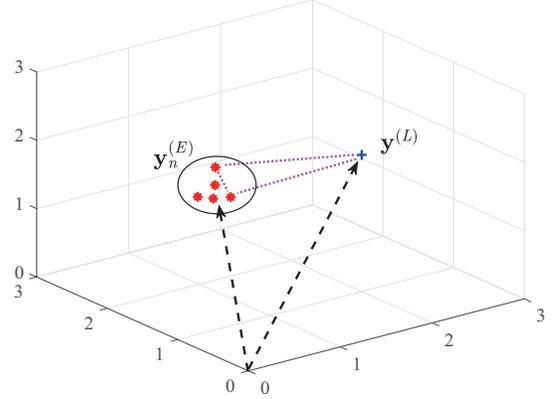}
	\caption{Basic principle of the proposed distance-based method.}
	\label{DistanceIntroduction} 
	\vspace{-5mm}
\end{figure}
In this case, we have the following two situations:
\begin{itemize}
\item $K>2$ and Eve hits LU's PS,
and we have
$\bm{y}^{(L)} = \bm{h}_L + \mathrm{e}^{\mathrm{j}\omega^{(L)}}\beta_{\mathrm{PSA}}^{(K)}\bm{h}_E + \bm{z}^{(L)}\in \Psi_{\mathrm{PSA}}^{(C)}$.

\item $K\geq2$ but Eve fails to hit LU's PS,
and we have $\bm{y}^{(L)} =  \bm{h}_L + \bm{z}^{(L)}\in \Psi_{\mathrm{PSA}}^{(C)}$.
\end{itemize}
Denote the channel observations in $\Psi_{\mathrm{PSA}}^{(C)}$ as $\{\bm{y}_1,\cdots,\bm{y}_{Q_C}\}$.
For both cases listed above, the BS needs to figure out which channel observation is $\bm{y}^{(L)}$.
We propose an unified method to handle both of the two cases, regardless the value of $K$.
Note that we have $\Psi_{\mathrm{PSA}}^{(C)}=\{\bm{y}^{(L)}\}\cup\Psi_{\mathrm{PSA}}^{(E)}$. From \eqref{ObservationSetPSACase1}, we observe that for a given realization of $\bm{h}_E$, the channel observations in $\Psi_{\mathrm{PSA}}^{(E)}$, after some phase shifts, are clustered around $\beta_{\mathrm{PSA}}^{(K)}\bm{h}_E$ in a $M${\it{--dimensional}} complex-valued space with relatively small biases due to the noise terms, i.e., $\bm{z}_1^{(E)},\bm{z}_2^{(E)},\cdots,\bm{z}_{Q_E}^{(E)}$. However, $\bm{y}^{(L)}$ is generally apart from $\beta_{\mathrm{PSA}}^{(K)}\bm{h}_E$ due to the existence of  $\bm{h}_L$. This observation inspires us a novel distance-based method to determine $\bm{y}^{(L)}$, the basic principle of which is shown in Fig \ref{DistanceIntroduction}.

\subsubsection{The distance-based method} For $\forall n\in \{0,1,\cdots,Q_C\}$, define the distances between the channel observations as
\begin{align}
\label{distanceInPSA}
d_{n}& \triangleq \mathbb{I}_{\{1\leq n\leq Q_C-1\}}\left(
\mathop{\mathrm{min}} \limits_{\phi}\left\|\bm{y}_n-\mathrm{e}^{\mathrm{j}\phi}\bm{y}_{n+1}\right\|^2\right)\nonumber \\
&\quad+
\mathbb{I}_{\{n=0\text{~or~}Q_C\}}\left(\mathop{\mathrm{min}}
\limits_{\phi}\left\|\bm{y}_{1}-\mathrm{e}^{\mathrm{j}\phi}\bm{y}_{Q_C}\right\|^2\right).
\end{align}
If $d_{n} > \epsilon$ and $d_{n-1} > \epsilon$,
then we conclude that $\bm{y}_{n}=\bm{y}^{(L)}$, and
if $d_{n} \leq \epsilon$ and $d_{n-1} \leq \epsilon$, then we conclude that $\bm{y}_n\in \Psi_{\mathrm{PSA}}^{(E)}$, where $\epsilon$ is a pre-designed threshold.
Note that it is possible that $d_{n} > \epsilon$ ($d_{n} < \epsilon$) while both $d_{n-1}$ and $d_{n+1}$ are smaller (larger) than $\epsilon$, which leads to a dilemma.
Therefore, a proper threshold $\epsilon$ should be designed to keep the probabilities of these events low.

\subsubsection{Design the threshold $\epsilon$}
If $\{\bm{y}_n,\bm{y}_{n+1}\}\subset\Psi_{\mathrm{PSA}}^{(E)}$, we expect the probability that $d_n$ exceeds $\epsilon$ to be as small as possible. Therefore, we choose the value of $\epsilon$ such that
\begin{align}
\mathcal{P}_F\left(\epsilon\right)\triangleq\mathcal{P}\left\{d_n > \epsilon | \{\bm{y}_n,\bm{y}_{n+1}\}\subset\Psi_{\mathrm{PSA}}^{(E)}\right\} \leq \eta,
\label{DeterminThreshold}
\end{align}
where $\eta$ ($0<\eta<1$) is a pre-designed value. In fact, $\eta$ is analogous to the false alarm rate in a hypothesis test problem, and in general, we should set a small value of $\eta$ to ensure a low EDR. Note that $\mathcal{P}\left\{d_n < \epsilon | \bm{y}_n=\bm{y}^{(L)}\text{~or~}\bm{y}_{n+1}=\bm{y}^{(L)}\right\}$ increases with $\epsilon$, and thus, we set $\mathcal{P}_F\left(\epsilon\right) = \eta$ to obtain the smallest decision threshold $\epsilon$ that satisfies \eqref{DeterminThreshold}.

The direct calculation of  $\epsilon$ via
 $\mathcal{P}_F\left(\epsilon\right)$ is difficult because there is no closed-form expression for  $\mathcal{P}_F\left(\epsilon\right)$.
To deal with this problem, in the following theorem, we provide an upper bound on $d_n$ when $\{\bm{y}_n,\bm{y}_{n+1}\}\subset\Psi_{\mathrm{PSA}}^{(E)}$ to obtain a closed-form approximation of $\mathcal{P}_F\left(\epsilon\right)$.
\begin{theorem}
\label{DecisionThreshold1}
A closed-form upper bound on $\mathcal{P}_F\left(\epsilon\right)$ can be written as
\begin{align}
\mathcal{P}_F\left(\epsilon\right)\leq\tilde{\mathcal{P}}_F\left(\epsilon\right)
\triangleq\Gamma\left(M,\epsilon/(2\sigma_z^2)\right)/\Gamma\left(M\right).
\label{DecisionThreshold1Upp}
\end{align}
\end{theorem}
\begin{IEEEproof}
Conditioned on $\{\bm{y}_n,\bm{y}_{n+1}\}\subset\Psi_{\mathrm{PSA}}^{(E)}$, we have
\begin{align}
d_{n} &= \mathop{\mathrm{min}}\limits_{\phi}
\Big\|\mathrm{e}^{\mathrm{j}\omega_{n}}\beta_{\mathrm{PSA}}^{(K)}\bm{h}_E + \bm{z}_{n}\nonumber \\
&\quad \quad \quad\quad -\mathrm{e}^{\mathrm{j}\phi}\left(\mathrm{e}^{\mathrm{j}\omega_{n+1}}\beta_{\mathrm{PSA}}^{(K)}
\bm{h}_E + \bm{z}_{n+1}\right)\Big\|^2 \nonumber \\
& \leq \tilde{d}_{n}  \triangleq\left\|\bm{z}_{n} - \mathrm{e}^{\mathrm{j}\left(\omega_{n}-\omega_{n+1}\right)}\bm{z}_{n+1}\right\|^2.
\label{dnApproximate}
\end{align}
In \eqref{dnApproximate}, $\left(\bm{z}_{n} - \mathrm{e}^{\mathrm{j}\left(\omega_{n}-\omega_{n+1}\right)}\bm{z}_{n+1}\right) \sim \mathbb{CN}\left(\bm{0},2\sigma_z^2\bm{I}\right)$, and thus $\tilde{d}_{n}\sim \mathbb{G}\left(M,2\sigma_z^2\right)$. According to \eqref{dnApproximate}, the CCDF of $\tilde{d}_{n}$, i.e., $\tilde{\mathcal{P}}_F\left(\epsilon\right)$ in \eqref{DecisionThreshold1Upp}, is an upper bound on $\mathcal{P}_F\left(\epsilon\right)$.
\end{IEEEproof}
Using $\tilde{d}_{n}$ in \eqref{dnApproximate} to approximate $d_n$ leads to an approximation of $\epsilon$, i.e.,
$\tilde{\epsilon} \triangleq 2\sigma_z^2\Gamma^{-1}\left(M,\eta\Gamma\left(M\right)\right)$,
where $\Gamma^{-1}\left(M,x\right)$ is the inverse function of $\Gamma\left(M,x\right)$.
We show the high accuracy of using $\tilde{d}_{n}$ to approximate $d_{n}$ in Fig. \ref{EDR:sub3} in Section \ref{Sectionpart}.
\subsubsection{EDR analysis}
Without loss of generality, assume $\bm{y}_1=\bm{y}^{(L)}$, then the BS may make a mistake on determining  $\bm{y}^{(L)}$ if one of the following events happens:
(1) $d_1\leq\epsilon$, denoted by $\mathcal{E}_1$,
(2) $d_0\leq\epsilon$, denoted by $\mathcal{E}_2$, and
(3) $\exists k\in \{2,3,\cdots,Q_C-1\}$ satisfying $d_{k}>\epsilon$, denoted by $\mathcal{E}_3$.
Accordingly, the EDR can be written as
$\mathcal{P}_{\mathrm{EDR}}\triangleq\mathcal{P}\left\{\bigcup_{i=1}^3\mathcal{E}_i\right\}.$

In general, a tractable expression for $\mathcal{P}_{\mathrm{EDR}}$ is hard to obtain due to the fact that: (1) the PDFs of $d_n$ for $n\in \left\{1,2,\cdots,Q_C\right\}$ are complicated, and (2) $\left\{d_0,d_1,\cdots,d_{Q_C}\right\}$ are correlated with each other. Here we consider an upper approximation of $\mathcal{P}_{\mathrm{EDR}}$, which is given by
\begin{align}
\mathcal{P}_{\mathrm{EDR}} &= \mathcal{P}\left\{\cup_{i=1}^3\mathcal{E}_i\right\}\leq
\mathcal{P}\left\{\mathcal{E}_1\right\}
+\mathcal{P}\left\{\mathcal{E}_2\right\}
+\mathcal{P}\left\{\mathcal{E}_3\right\}\nonumber \\
&\approx  \overline{\mathcal{P}}_{\mathrm{EDR}} \triangleq 2\mathcal{P}\left\{\mathcal{E}_1\right\}+\eta,
\label{PedrUpperapproximation}
\end{align}
where we approximate $\mathcal{P}\left\{\mathcal{E}_3\right\}$ as $\eta$
based on the observation that we have $\mathcal{P}\{d_n>\epsilon\}\leq\eta$  for $\forall n\in\left\{2,3,\cdots,Q_C-1\right\}$, and when $\eta$ is small, the probability of having multiple elements in $\left\{d_2,d_3,\cdots,d_{Q_C-1}\right\}$
larger than $\epsilon$ is negligible.

Now, to calculate $\overline{\mathcal{P}}_{\mathrm{EDR}}$ is equivalent to calculate $\mathcal{P}\left\{d_1\leq \epsilon\right\}$. However, since there is no  tractable expression for the PDF of $d_1$, it is still very difficult if not impossible.
To deal with this problem, we provide a tractable approximate of $d_1$ as follows:
\begin{itemize}
\item if $\bm{y}_1$ is contaminated by $\bm{h}_E$, we denote $d_1$ in this case as $d_1^{(C)}$, and then we have
\begin{align}
&d_{1}^{(C)} \triangleq \mathop{\mathrm{min}} \limits_{\phi} \Big\| \bm{h}_L + \mathrm{e}^{\mathrm{j}\omega_1}\beta_{\mathrm{PSA}}^{(K)}\bm{h}_E + \bm{z}_1 \nonumber \\
& \quad\quad\quad\quad\quad - \mathrm{e}^{\mathrm{j}\phi}\left(\mathrm{e}^{\mathrm{j}\omega_2}\beta_{\mathrm{PSA}}^{(K)}\bm{h}_E + \bm{z}_2\right)\Big\|^2 \nonumber \\
\approx \ &\tilde{d}_{1}^{(C)}
\triangleq
\mathop{\mathrm{min}} \limits_{\phi} \Big\|\bm{h}_L + \bm{z}_1 - \tilde{\bm{z}}_2
 \nonumber \\
&\quad\quad\quad\quad\quad + \left(\mathrm{e}^{\mathrm{j}\omega_1}-
\mathrm{e}^{\mathrm{j}\phi}\mathrm{e}^{\mathrm{j}\omega_2}\right)\beta_{\mathrm{PSA}}^{(K)}\bm{h}_E
\Big\|^2,
\label{dC}
\end{align}
\item if $\bm{y}_1$ is not contaminated by $\bm{h}_E$, we denote $d_1$ in this case as $d_1^{(N)}$, and then we have
\begin{align}
&d_{1}^{(N)}\triangleq
\mathop{\mathrm{min}} \limits_{\phi} \left\|\bm{h}_L + \bm{z}_1 - \mathrm{e}^{\mathrm{j}\phi}\left(\mathrm{e}^{\mathrm{j}\omega_2}\beta_{\mathrm{PSA}}^{(K)}\bm{h}_E + \bm{z}_{2}\right)\right\|^2\nonumber\\
\approx\ &\tilde{d}_{1}^{(N)}\triangleq
\mathop{\mathrm{min}} \limits_{\phi} \left\|\bm{h}_L + \bm{z}_1 - \tilde{\bm{z}}_2  - \mathrm{e}^{\mathrm{j}\phi}\mathrm{e}^{\mathrm{j}\omega_2}\beta_{\mathrm{PSA}}^{(K)}\bm{h}_E
\right\|^2,
\label{dN}
\end{align}
\end{itemize}
where the approximations in \eqref{dC} and \eqref{dN} is obtained by replacing $\mathrm{e}^{\mathrm{j}\phi}\pmb{z}_2$ with $\tilde{\pmb{z}}_2\sim\mathbb{CN}\left(\pmb{0},\sigma_z^2\pmb{I}_M\right)$ which is independent of $\phi$.
Both of $\tilde{d}_{1}^{(C)}$ and $\tilde{d}_{1}^{(N)}$ are further lower bounded by
\begin{align}
\tilde{d}_{1}^{(i)}&\geq
\mathop{\mathrm{min}} \limits_{x} \left\|\bm{h}_L + \bm{z}- x\bm{h}_E
\right\|^2\nonumber\\
&=
\left\|\bm{h}_L + \bm{z}\right\|^2-\frac{\left|\bm{h}_E^H\left(\bm{h}_L + \bm{z} \right)\right|^2}{\left\|\bm{h}_E\right\|^2} \nonumber \\
&\overset{(a)}{\approx}\hat{d}_1\triangleq \left(\bm{h}_L + \bm{z} \right)^H\bm{Q}\left(\bm{h}_L + \bm{z}\right) , \quad i\in\{C,N\},
\label{d1LowerBound}
\end{align}
where $\bm{z}\triangleq\bm{z}_1-\tilde{\bm{z}}_2$, $\bm{Q}\triangleq \bm{I} - \frac{\bm{R}_{E}}{\mathrm{Tr}\left(\bm{R}_{E}\right)}$, and step $(a)$ is obtained by using $\frac{\bm{h}_E\bm{h}_E^H}{\left\|\bm{h}_E\right\|^2}\rightarrow \frac{\bm{R}_{E}}{\mathrm{Tr}\left(\bm{R}_{E}\right)}$ when $M$ is large.

According to \eqref{d1LowerBound}, we can approximate $\mathcal{P}\left\{\mathcal{E}_1\right\}$ by $\mathcal{P}\left\{\hat{d}_1\leq\epsilon\right\}$, and the calculation of $\mathcal{P}\left\{\hat{d}_1\leq\epsilon\right\}$ is provided in the following theorem.
\begin{theorem}
\label{Theorem:d1lower}
$\mathcal{P}\left\{\hat{d}_1\leq\epsilon\right\}$ can be calculated as
\begin{align}
\mathcal{P}\left\{\hat{d}_1\leq\epsilon\right\}=
\mathcal{P}\left\{ \bm{\alpha}^H\bar{\bm{Q}}\bm{\alpha}\leq\epsilon\right\},
\end{align}
where $\bm{\alpha}\sim\mathbb{CN}\left(\bm{0},\bm{I}_M\right)$, and
$\bar{\bm{Q}} \triangleq \left(\bm{R}_{L}+2\sigma_z^2\pmb{I}_M\right)^{\frac{1}{2}}
\bm{Q} \left(\bm{R}_{L}+2\sigma_z^2\pmb{I}_M\right)^{\frac{1}{2}}$.
\end{theorem}
\begin{IEEEproof}
The proof  is similar to that of  Theorem \ref{Theorem:T12}, and thus is omitted.
\end{IEEEproof}
We plot the CCDF of $d_1^{(C)}$,$d_1^{(N)}$, and $\hat{d}_1$ in Fig. \ref{EDR:sub3} to show that $d_1^{(N)}$ is a satisfying lower approximation of both $d_1^{(C)}$ and $d_1^{(N)}$.
And we also compare $\mathcal{P}_{\mathrm{EDR}}$ obtained from simulation with its analytical upper approximation $\overline{\mathcal{P}}_{\mathrm{EDR}}$ in \eqref{PedrUpperapproximation} in Section VII.

\subsection{If $K$ is unknown by the BS}
We have discussed the cases when $K$ is known by the BS in the previous subsections. In practice, $K$ is chosen by Eve and the BS generally does not have any prior information on $K$. However, even though the BS does not know the value of $K$, we will show that in order to identify $\bm{y}^{(L)}$, the BS only needs to carry out some additional operations to decide whether a hit event happens. And then, all the discussions in the previous subsections hold as well.
In the following, the detailed discussions on these additional operations are provided. For simplicity and clarity, we classify the discussions according to the value of $Q_C$.
\subsubsection{$Q_C=1$}
In this case, $\Psi_{\mathrm{PSA}}^{(C)}=\left\{\bm{y}^{(L)}\right\}$, and
the BS needs to distinguish between:
\begin{itemize}
    \item $\mathcal{H}^{(0,1)}$: $K=0$, i.e., Eve keeps silent,
    \item $\mathcal{H}^{(1,1)}$: $K=1$ and Eve successfully hits LU's PS.
\end{itemize}
Note that to distinguish between $\mathcal{H}^{(0,1)}$ and $\mathcal{H}^{(1,1)}$ is equivalent to detect the existence of PSA under the conventional channel training scheme wherein only one PS is assigned to the LU as discussed in \cite{J.K.Tugnait,J.K.Tugnait2015WCL,
J.M.Kang2015VTC,Q.Xiong2015TIFS,D.Kapetanovic2013PIMRC}.
And this can be easily realized by checking the LLR, i.e.,
\begin{align}
\label{OriginalDetection}
\ln p\left\{\bm{y}^{(L)}|\mathcal{H}^{(0,1)}\right\}
-\ln p\left\{\bm{y}^{(L)}|\mathcal{H}^{(1,1)}\right\}
{\gtreqless}_{\mathcal{H}^{(1,1)}}^{\mathcal{H}^{(0,1)}} 0,
\end{align}
where $\bm{y}^{(L)}\sim\mathbb{CN}\left(\bm{0},\bm{R}_{L,z}\right)$ conditioned on $\mathcal{H}_0$, and $\bm{y}^{(L)}\sim\mathbb{CN}\left(\bm{0},\bm{R}_{L,E,z}^{(1)}\right)$ conditioned on $\mathcal{H}_1$.

\subsubsection{$Q_C=2$}
The following two situations will result in $Q_C=2$, i.e.,
\begin{itemize}
    \item $\mathcal{H}^{(1,2)}$: $K=1$, but Eve fails to hit LU's PS,
    \item $\mathcal{H}^{(2,2)}$: $K=2$, and Eve successfully hits LU's PS.
\end{itemize}
To distinguish between $\mathcal{H}^{(1,2)}$ and $\mathcal{H}^{(2,2)}$, we first determine the subcases of $\mathcal{H}^{(1,2)}$, i.e., $\mathcal{H}_0^{(1,2)}$ and $\mathcal{H}_1^{(1,2)}$ defined in Section \ref{K1FailHit}, by using the method  provided in Section \ref{K1FailHit}. Assume the decision result between $\mathcal{H}_0^{(1,2)}$ and $\mathcal{H}_1^{(1,2)}$ is denoted by $\mathcal{H}_{j_1}^{(1,2)}$ where $j_1=\mathrm{argmax}_{i=0,1}~p\left( \bm{g} |\mathcal{H}_{i}^{(1,2)}\right)$.
Then, we determine the subcases of $\mathcal{H}^{(2,2)}$, i.e., $\mathcal{H}_0^{(2,2)}$ and $\mathcal{H}_1^{(2,2)}$ defined in  Section \ref{Expample}, by using the method  provided in Section \ref{Expample} (both the power comparison based method and the GLLR based method are applicable).
If the GLLR based method is used, then we denote the decision result between $\mathcal{H}_0^{(2,2)}$ and $\mathcal{H}_1^{(2,2)}$ as $\mathcal{H}_{j_2}^{(2,2)}$, where $j_2=\mathrm{argmax}_{i=0,1} \left(\mathrm{max}_{\omega^{(L)},\omega^{(E)}}~p\left( \bm{g} |\mathcal{H}_{i}^{(2,2)}\right)\right)$.
After obtaining $\mathcal{H}_{j_1}^{(1,2)}$ and $\mathcal{H}_{j_2}^{(2,2)}$, we now resort to
GLLR test to determine which one of  $\mathcal{H}^{(1,2)}$ and $\mathcal{H}^{(2,2)}$ is true, i.e.,
$\ln \left(\mathrm{max}_{\omega^{(L)},\omega^{(E)}}~p\left(\bm{g}|\mathcal{H}_{j_2}^{(2,2)}\right)\right)
-\ln p\left(\bm{g}|\mathcal{H}_{j_1}^{(1,2)}\right)
\gtreqless_{\mathcal{H}^{(1,2)}}^{\mathcal{H}^{(2,2)}} 0$.
\subsubsection{$Q_C>2$}
In this cases, we can first recognize which channel observation in $\Psi_{\mathrm{PSA}}^{(C)}$ is $\bm{y}^{(L)}$ by using the distance-based method proposed in Section \ref{Section:ConterAttack:LargeL}.
After obtaining $\bm{y}^{(L)}$, we only need to determine whether Eve hits LU's PS, which can be simply realized by checking the LLR which is similar to that in \eqref{OriginalDetection}.

Combining all these steps in the discussions above, we can determine which PS is $\bm{x}^{(L)}$ (which channel observation is $\bm{y}^{(L)}$), even when $K$ is unknown by the BS.

\section{Channel Estimation under PSA}
\label{ChannelEsPSA}
After determining which PS is $\bm{x}^{(L)}$, now the BS can estimate both of the legitimate and illegitimate channels.
\subsection{Estimation of the legitimate channel}
Assume minimum mean square error (MMSE) estimator is used at the BS, then the estimation of LU's channel and corresponding the mean square error (MSE) matrices are given by
\begin{align}
\hat{\bm{h}}_L&=\mathbb{I}_{\{\bm{x}^{(L)}\in\Phi_E\}}\bm{R}_L\left(\bm{R}_{L,E,z}^{(K)} \right)^{-1}\bm{y}^{(L)}\nonumber \\
&\quad+
\mathbb{I}_{\{\bm{x}^{(L)}\notin\Phi_E\}}\bm{R}_L\left(\bm{R}_{L,z} \right)^{-1}\bm{y}^{(L)},\\
\tilde{\bm{R}}_L&=\mathbb{E}\left\{\left(\bm{h}_L-\hat{\bm{h}}_L\right)
\left(\bm{h}_L-\hat{\bm{h}}_L\right)^H\right\}\nonumber\\
&=\mathbb{I}_{\{\bm{x}^{(L)}\in\Phi_E\}}\left(\bm{R}_L-\bm{R}_L\left(\bm{R}_{L,E,z}^{(K)} \right)^{-1}\bm{R}_L\right)
\nonumber \\
&\quad +\mathbb{I}_{\{\bm{x}^{(L)}\notin\Phi_E\}}\left(\bm{R}_L-\bm{R}_L\left(\bm{R}_{L,z} \right)^{-1}\bm{R}_L\right).
\end{align}
\subsection{Estimation of illegitimate channel}
To estimate the channel of Eve, we consider the following two situations:
\subsubsection{$Q_C=1$ and PSA is detected} In this case, the only one channel observation in $\Psi_C$ is $\bm{y}^{(L)} = \bm{h}_L+\mathrm{e}^{\mathrm{j}\omega^{(L)}}\beta_{\mathrm{PSA}}^{(1)}\bm{h}_E+\bm{z}^{(L)}$, and therefore, the estimation of Eve's channel\footnote{We have to point out here that the estimated illegitimate channel, i.e., $\hat{\bm{h}}_E$, is, in fact, not an estimation of $\bm{h}_E$ but an estimation of the entirety of the product of $\mathrm{e}^{\mathrm{j\omega^{(L)}}}$ and $\bm{h}_E$, i.e., $\mathrm{e}^{\mathrm{j}\omega^{(L)}}\bm{h}_E$. Due to the fact that $\bm{h}_E$ is a  circular symmetrical complex random vector, $\mathrm{e}^{\mathrm{j}\omega^{(L)}}\bm{h}_E$ has the same distribution as $\bm{h}_E$, even though $\omega^{(L)}$ is unknown.} and the corresponding MSE matrix are given by
\begin{align}
\label{channelMSE}
\hat{\bm{h}}_E&= \beta_{\mathrm{PSA}}^{(1)}\bm{R}_E\left(\bm{R}_{L,E,z}^{(1)}\right)^{-1}\bm{y}^{(L)},\\
\tilde{\bm{R}}_E&=\mathbb{E}\left\{\left(\mathrm{e}^{\mathrm{j}\omega^{(L)}}\bm{h}_E-\hat{\bm{h}}_E\right)
\left(\mathrm{e}^{\mathrm{j}\omega^{(L)}}\bm{h}_E-\hat{\bm{h}}_E\right)^H\right\}\nonumber\\
&=\bm{R}_E-\left|\beta_{\mathrm{PSA}}^{(1)}\right|^2
\bm{R}_E\left(\bm{R}_{L,E,z}^{(1)}\right)^{-1}\bm{R}_E.
\end{align}

\subsubsection{$Q_C > 1$} In this case, we can estimate $\bm{h}_E$ from the channel observations in $\Psi_{\mathrm{PSA}}^{(E)}$. Denote the channel observations in $\Psi_{\mathrm{PSA}}^{(E)}$ as $\bm{y}_i^{(E)}=\beta_{\mathrm{PSA}}^{(K)}\bm{h}_E^{(i)} + \bm{z}_i^{(E)}$, for $i = 1,2,\cdots,Q_E$, where we have $\bm{h}_E^{(i)}\triangleq\mathrm{e}^{\mathrm{j}\omega_i^{(E)}}\bm{h}_E$.
To estimate the illegitimate channel, we first combine $\bm{y}_1^{(E)},\bm{y}_2^{(E)},\cdots,\bm{y}_{Q_E}^{(E)}$  as
\begin{align}
\bm{y}_{E} \triangleq \frac{1}{Q_E}\sum_{i=1}^{Q_E}\kappa_i\bm{y}_{i}^{(E)}= \frac{1}{Q_E}\sum_{i=1}^{Q_E} \kappa_i\beta_{\mathrm{PSA}}^{(K)}\bm{h}_E^{(i)} + \tilde{\bm{z}}_E,
\label{EstimationofHEHighnoise}
\end{align}
where $\kappa_i\triangleq \langle\bm{y}_{1}^{(E)},\bm{y}_{i}^{(E)} \rangle\big/|\langle\bm{y}_{1}^{(E)},\bm{y}_{i}^{(E)}\rangle|$, for $i=1,2,\cdots,Q_E$,
are the combination coefficients, and $\tilde{\bm{z}}_E\triangleq\frac{1}{Q_E}\sum_{i=1}^{Q_E}\kappa_i \bm{z}_{i}^{(E)}$ is the noise term.
Note that the exact distribution of $\tilde{\bm{z}}_E$, or even its covariance matrix, is very hard to obtain because the combination coefficients, i.e., $\{\kappa_i\}_{i=1}^{Q_E}$, are determined by $\{\bm{y}_i^{(E)}\}_{i=1}^{Q_E}$ in a extremely complicated form. However, we observe that the noise floor of $\bm{z}_{i}^{(E)}$ decreases with the increase of $\tau$ and $p_L$, and when $\tau$ and $p_L$ is sufficiently large, it reasonable to approximate $\kappa_i$ as
$\kappa_i\approx\tilde{\kappa}_i\triangleq\mathrm{e}^{\mathrm{j}\left(\omega_1^{(E)}-\omega_i^{(E)}\right)}$.
Based on $\tilde{\kappa}_i$, we have $\mathbb{E}\left\{\bm{z}_{i}^{(E)}\left(\bm{z}_{i}^{(E)}\right)^H\right\}
\approx\frac{\sigma_z^2}{Q_E}\bm{I}_M$.
Inserting $\kappa_i\approx\tilde{\kappa}_i$ into \eqref{EstimationofHEHighnoise}, we obtain $\bm{y}_{E}\approx \beta_{\mathrm{PSA}}^{(K)}\bm{h}_E^{(1)} + \tilde{\bm{z}}_E$.
Using the linear MMSE estimator to estimate the illegitimate channel, we have
\begin{align}
\hat{\bm{h}}_E &= \beta_{\mathrm{PSA}}^{(K)}\bm{R}_E\left(\bm{R}_{E}^{(K)} + \frac{\sigma_z^2}{Q_E}\bm{I}_M\right)^{-1}\bm{y}_{E},\\
\tilde{\bm{R}}_E&=\bm{R}_E-\left|\beta_\mathrm{PSA}^{(K)}\right|^2\bm{R}_E\left(\bm{R}_{E}^{(K)} + \frac{\sigma_z^2}{Q_E}\bm{I}_M\right)^{-1}\bm{R}_E.
\end{align}

\section{Secure Transmission under PSA}
\label{PSASecureTransmission}
After UCTP, based on the channel estimations, SB vector is designed to transmit the confidential messages. Denote
the signals received by the LU and Eve during DDTP as $y_L$ and $y_E$, respectively, then $y_q$ for $q\in\{L,E\}$ can be written as
\begin{align}
y_q &= \sqrt{p_B}\left(\hat{\bm{h}}_q+\tilde{\bm{h}}_q\right)^H\bm{v}x + n_q,\label{ReceivedSignalLU}
\end{align}
where $\hat{\bm{h}}_q$ for $q\in\{L,E\}$ are the estimations of the legitimate and illegitimate channels, respectively,
$\tilde{\bm{h}}_q\triangleq\bm{h}_q-\hat{\bm{h}}_q$ for $q\in\{L,E\}$ are the corresponding  estimation errors,
$n_q\sim\mathbb{CN}\left(0,\sigma_q^2\right)$ for $q\in\{L,E\}$ are the additive white Gaussian noises, $\bm{v}$ satisfying $\left\|\bm{v}\right\|=1$ is the SB vector designed by the BS, $x\sim\mathbb{CN}\left(0,1\right)$ is the confidential signal, and $p_B$ is the power budget of the BS.
Note that if positive secrecy rate exists, maximum power is optimal \cite{S.Loyka2012ISITP}.

We make an reasonable assumption here that the LU and Eve can know their equivalent channel coefficients, i.e., $\left(\hat{\bm{h}}_L+\tilde{\bm{h}}_L\right)^H\bm{v}$ and $\left(\hat{\bm{h}}_E+\tilde{\bm{h}}_E\right)^H\bm{v}$, when they decode the information-bearing symbols.
This assumption can be realized in the following two ways,
\begin{itemize}
    \item after obtaining the beamforming vector $\bm{v}$, the BS performs a downlink training procedure to let the LU estimate its equivalent channel coefficient.
    \item note that the equivalent channel is only a complex number, and the LU may direct learn it through the received signal symbols during the whole DDTP.
\end{itemize}
Based on this assumption, the $\mathrm{SNR}$ of the LU and Eve can be written, respectively, as
$
\mathrm{SNR}_L = \frac{p_B\left|\bm{v}^H\left(\hat{\bm{h}}_L+\tilde{\bm{h}}_L\right)\right|^2}{\sigma_L^2}$ and
$\mathrm{SNR}_E = \frac{p_B\left|\bm{v}^H\left(\hat{\bm{h}}_E+\tilde{\bm{h}}_E\right)\right|^2}{\sigma_E^2}
$. Then
the secrecy rate is given by
\begin{align}
R_S = \ln\left(\frac{1 + \mathrm{SNR}_L}{1 + \mathrm{SNR}_E}\right)
=\ln\left(\frac{\bm{v}^H\bm{H}_L\bm{v}}
{\bm{v}^H\bm{H}_E\bm{v}}\right),
\label{Rate1}
\end{align}
where $\bm{H}_q\triangleq\bm{I}+\frac{p_{B}}{\sigma_q^2}\left(\hat{\bm{h}}_q+
\tilde{\bm{h}}_q\right)\left(\hat{\bm{h}}_q+
\tilde{\bm{h}}_q\right)^H$ for $q\in\{L,E\}$.
Note that $\tilde{\bm{h}}_L$ and $\tilde{\bm{h}}_E$ are the channel estimation errors which are unknown by the BS when designing $\bm{v}$, therefore, the BS is not able to maximize the instantaneous secrecy rate.
To secure the data transmission, the BS can alternatively maximize the \emph{average} secrecy rate over the unknown channel estimation errors $\tilde{\bm{h}}_L$ and $\tilde{\bm{h}}_E$, i.e.,
\begin{align}
\mathop{\mathrm{max}} \limits_{\bm{v}, \left\|\bm{v}\right\|^2=1} \quad
\mathbb{E}_{\tilde{\bm{h}}_B,\tilde{\bm{h}}_E}\left\{R_S\right\}.
\label{AverageRateMax}
\end{align}
In general, solving \eqref{AverageRateMax} is complicated because the expectation operation in \eqref{AverageRateMax} leads to an intractable expression of the objective.
To handle it, the following lemma provides us a tractable approximation of the objective in \eqref{AverageRateMax}.
\begin{lemma}
\label{Approximation1}
For two random variables $X$ and $Y$ satisfying $X,Y\geq0$, $\ln\left(\frac{1+\mathbb{E}_{X}\left\{X\right\}}{1+\mathbb{E}_{Y}\left\{Y\right\}}\right)$ can approximate $\mathbb{E}_{X,Y}\left\{\ln\left(\frac{1+X}{1+Y}\right)\right\}$ in the sense that they have common upper and lower bounds.
\end{lemma}
\begin{IEEEproof}
This lemma is inspired by \cite[Lemma 1]{Q.ZhangTJSTSP2014JSTSP}, and the proof is given in Appendix \ref{Lemma1Lemma2Proof}.
\end{IEEEproof}

Using Lemma \ref{Approximation1}, problem \eqref{AverageRateMax} is approximated by
\begin{align}
\label{secrecyrateApproximation}
\mathop{\mathrm{max}} \limits_{\bm{v}, \left\|\bm{v}\right\|^2=1}\quad
\frac{\bm{v}^H\bar{\bm{H}}_L\bm{v}
}
{\bm{v}^H\bar{\bm{H}}_E\bm{v}},
\end{align}
where we have $\bar{\bm{H}_q}\triangleq\bm{I}+\frac{p_{B}}{\sigma_E^2}\left(\hat{\bm{h}}_q
\hat{\bm{h}}_q^H + \tilde{\bm{R}}_q\right)$ for $q\in\{L,E\}$.
Note that similar approximations are also used in \cite{M.K.Member2007JSAC,X.Wang2013TVT} for mathematical tractability.
Problem \eqref{secrecyrateApproximation} is to maximize a generalized Rayleigh quotient, whose optimal solution is given by
$\bm{v}_{\mathrm{opt}} = \bm{u}_{\mathrm{max}}\left\{\left(\bar{\bm{H}}_E\right)^{-1}\left(\bar{\bm{H}}_L\right)\right\}$.
The calculation of $\bm{v}_{\mathrm{opt}}$ may be computationally complex when $M$ gets large. In fact, \eqref{secrecyrateApproximation} can be lower bounded by
\begin{align}
\label{AverageRateMax2}
\mathop{\mathrm{max}} \limits_{\tilde{\bm{v}}, \left\|\tilde{\bm{v}}\right\|^2=1} \quad
\frac{p_{B}}{\sigma_L^2}\frac{\tilde{\bm{v}}^H
\hat{\bm{h}}_L\hat{\bm{h}}_L^H
\tilde{\bm{v}}}
{\tilde{\bm{v}}^H\bar{\bm{H}}_E\tilde{\bm{v}}}.
\end{align}
and the optimal beamforming vector for \eqref{AverageRateMax2} is
$\tilde{\bm{v}}_{\mathrm{opt}} =\rho\left(\bm{H}_E\right)^{-1}\hat{\bm{h}}_L$.
where $\rho$ is a scaling factor chosen to satisfy $\tilde{\bm{v}}_{\mathrm{opt}}^H\tilde{\bm{v}}_{\mathrm{opt}}=1$.

\begin{remark}
The matrix inverse operation in $\bm{v}_{\mathrm{opt}}$ and $\tilde{\bm{v}}_{\mathrm{opt}}$ are in the form of $\left(\bm{B}+\bm{\mu}\bm{\mu}^H\right)^{-1}$, where $\bm{B}=\bm{I}+\frac{p_B}{\sigma_E^2}\tilde{\bm{R}}_E$ and $\bm{\mu} = \sqrt{\frac{p_B}{\sigma_E^2}}\hat{\bm{h}}_E$, which can be implemented by using the Matrix Inverse Lemma, i.e., $\left(\bm{B}+\bm{\mu}\bm{\mu}^H\right)^{-1}=\bm{B}^{-1} - \frac{\bm{B}^{-1}\bm{\mu}\bm{\mu}^H\bm{B}^{-1}}{1+\bm{\mu}^H\bm{B}^{-1}\bm{\mu}}$.
Note that $\tilde{\bm{R}}_E$ is the MSE matrix which is usually known previously, therefore, $\bm{B}^{-1}$ can be precalculated to reduce the real time computational complexity.
\end{remark}

\section{Secure Transmission under PJA}
\label{STJA}
In this section, we discuss the method to deal with the PJA during UCTP.
We follow the same steps as what we did to deal with PSA in previous sections, i.e., we first determine which PS is transmitted by the LU, and then we estimate the legitimate and illegitimate channels, and finally we design the secure beamforming vector.

\subsection{Determine which PS is transmitted by the LU}
We have to point out that
to identify $\bm{y}^{(L)}$ from $\Psi$  under PJA is more complicated than that under PSA.
This is because
under PSA, the amplitude of $\bm{h}_E$ projected on each PS, i.e., $\beta_{\mathrm{PSA}}^{(K)}$, is fixed, and only the phase shifts, i.e., $\{\omega_1^{(E)},\cdots,\omega_K^{(E)}\}$, are unknown. However, under PJA, both the amplitudes and the phase shifts are unknown because $\{\mu_1,\cdots,\mu_N\}$ are complex-valued random variables.
To make a decision, we further develop the distance-based method proposed in Section \ref{Section:ConterAttack:LargeL} to make it applicable to combat with the PJA.

For $n = {1,2,\cdots,N}$, we redefine the distance between two channel observations as
\begin{align}
\label{distanceInJA}
\left\{
\begin{aligned}
d_{\mathrm{PJA},n}^{\left(+\right)} &\triangleq
\mathbb{I}_{\{1\leq n<N\}}\left(\mathop{\mathrm{min}}\limits_{a} \left\|\bm{y}_n-a\bm{y}_{n+1}\right\|^2\right) \\
&\quad +
\mathbb{I}_{\{n=N\}}\left(\mathop{\mathrm{min}}\limits_{a} \left\|\bm{y}_N-a\bm{y}_{1}\right\|^2\right),\\
d_{\mathrm{PJA},n}^{\left(-\right)} &\triangleq
\mathbb{I}_{\{1< n\leq N\}}\left(\mathop{\mathrm{min}}\limits_{a} \left\|\bm{y}_n-a\bm{y}_{n-1}\right\|^2\right)  \\
&\quad+
\mathbb{I}_{\{n=1\}}\left(\mathop{\mathrm{min}}\limits_{a} \left\|\bm{y}_1-a\bm{y}_{N}\right\|^2\right).
\end{aligned}\right.
\end{align}
Note that \eqref{distanceInJA} is  different from \eqref{distanceInPSA} in the sense that the parameter $a$ is a general complex-valued number but not in the form of $\mathrm{e}^{\mathrm{j}\varphi}$. Besides, in general, we have $d_{\mathrm{PJA},n+1}^{\left(-\right)}\neq d_{\mathrm{PJA},n}^{\left(+\right)}$.
The decision method is modified as:
if $d_{\mathrm{PJA},n}^{(+)}\geq\epsilon$ and $d_{\mathrm{PJA},n+1}^{(-)}\geq\epsilon$, then we decide that $\bm{y}_n=\bm{y}^{(L)}$.

In general, a proper threshold $\epsilon$ should be designed to keep a low EDR. Therefore, we choose
$\epsilon$ such that it satisfies
\begin{align}
&\mathcal{P}\left\{d_{\mathrm{PJA},n}^{(+)}>\epsilon
|\{\bm{y}_n,\bm{y}_{n+1}\}\subset\Psi_{\mathrm{PJA}}^{(E)}\right\}
\nonumber \\
&\overset{(*)}{=}
\mathcal{P}\left\{d_{\mathrm{PJA},n+1}^{(-)}>\epsilon
|\{\bm{y}_n,\bm{y}_{n+1}\}\subset\Psi_{\mathrm{PJA}}^{(E)}\right\}\leq\eta,
\end{align}
where step $(*)$ is because $d_{\mathrm{PJA},n}^{(+)}$ has the same distribution as $d_{\mathrm{PJA},n+1}^{(-)}$, if $\{\bm{y}_n,\bm{y}_{n+1}\}\subset\Psi_{\mathrm{PJA}}^{(E)}$.
Note that the exact distribution of $d_{\mathrm{PJA},n}^{(+)}$ is hard to obtain, and we provide a closed-form upper bound on  $\mathcal{P}\left\{d_{\mathrm{PJA},n}^{(+)}>\epsilon
|\{\bm{y}_n,\bm{y}_{n+1}\}\subset\Psi_{\mathrm{PJA}}^{(E)}\right\}$ as an approximation in the following theorem.
\begin{theorem}
\label{JammingAttackUpperbound}
If $\{\bm{y}_n,\bm{y}_{n+1}\}\subset\Psi_{\mathrm{PJA}}^{(E)}$, then $\mathcal{P}\left\{d_{\mathrm{PJA},n}^{(+)}>\epsilon\right\}$ is upper bounded by
\begin{align}
\mathcal{P}_{\mathrm{PJA}}^{(\mathrm{F})}\left(\epsilon\right)
&\triangleq\frac{\sigma_z^2M}{\epsilon}-\mathrm{e}^{-\frac{\epsilon}{\sigma_z^2}}\sum_{m=0}^{M-1}\sum_{k=0}^m\frac{1}{k!}\left(\frac{\epsilon }{\sigma_z^2}\right)^{k-1}.
\label{DJAapproximate}
\end{align}
\end{theorem}
\begin{IEEEproof}
Please refer to Appendix \ref{Appendix:ProofUpperBound}.
\end{IEEEproof}
It can be easily verified that $\mathcal{P}_{\mathrm{PJA}}^{(\mathrm{F})}\left(\epsilon\right)$ decreases with $\epsilon$. Therefore, we can use the bisection method to search $\tilde{\epsilon}$ such that
$\mathcal{P}_{\mathrm{PJA}}^{(\mathrm{F})}\left(\tilde{\epsilon}\right)=\eta$, and use $\tilde{\epsilon}$ as the decision threshold.

Under PJA, the distributions of $d_{\mathrm{PJA},n}^{(+)}$ and $d_{\mathrm{PJA},n}^{(-)}$ are extremely complicated if $\bm{y}_n=\bm{y}^{(L)}$, and therefore the EDR analysis is left for future work.
Here, we have to emphasize that the proposed distance-based method only relies on the difference between $\bm{h}_L$ and $\bm{h}_E$, but not the difference between $\bm{R}_L$ and $\bm{R}_E$. Therefore, even if $\bm{R}_L=\bm{R}_E$, the proposed method still works.

\subsection{Channel Estimation}
Without loss of generality, in this subsection, we assume $\bm{y}_1=\bm{y}^{(L)}$.
We can use linear MMSE estimator to estimate the legitimate channel, i.e.,
$\hat{\bm{h}}_L=\bm{R}_L\left(\bm{R}_L+\frac{1}{\tau}\left|\beta_{\mathrm{PJA}}\right|^2\bm{R}_E + \sigma_z^2\bm{I}\right)^{-1}\bm{y}_1$,
and thus, the corresponding MSE matrix becomes
$\tilde{\bm{R}}_{\mathrm{PJA},L}=
\bm{R}_L - \bm{R}_L\left(\bm{R}_L+\frac{1}{\tau}\left|\beta_{\mathrm{PJA}}\right|^2\bm{R}_E + \sigma_z^2\bm{I}\right)^{-1}\bm{R}_L$.
Note that under PJA, the linear MMSE estimator is not the optimal estimator in the MMSE sense due to the fact that $u_1\bm{h}_E$ is not a Gaussian random vector.
As for the estimation of $\bm{h}_E$, we have to point out that though
$\bm{y}_2,\bm{y}_3,\cdots,\bm{y}_N$ contain $\bm{h}_E$, it is still hard to estimate $\bm{h}_E$ due to the unknown parameters $\mu_k$ for $k=2,3,\cdots,N$.
Define $\bm{Y}_E\triangleq\left[\bm{y}_2,\bm{y}_3,\cdots,\bm{y}_N\right]$, we propose to only estimate the directional information of $\bm{h}_E$ in a $M${\it{--dimensional}} complex-valued vector space as,
\begin{align}
\label{DirectionEstimation}
\check{\bm{h}}_E = {\mathrm{argmax}}_{\left\|\bm{x}\right\|^2=1}\quad\bm{x}^H\bm{Y}_E \bm{Y}_E^H\bm{x},
\end{align}
and obviously, we get $\check{\bm{h}}_E = \bm{u}_{\mathrm{max}}\left(\bm{Y}_E\bm{Y}_E^H\right)$.
The principle behind the estimator in \eqref{DirectionEstimation} can be summarized as follows.
Each column of the matrix $\bm{Y}_E$ is in the form of $\mu\bm{h}_E + \bm{z}$. The value of $\bm{z}$ is generally small, and therefore, the characteristic space of $\bm{Y}_E\bm{Y}_E^H$ should be dominated by $\bm{h}_E$.

\subsection{Secure Transmission}
Note that only the directional information of the illegitimate channel, i.e, $\check{\bm{h}}_E$, is obtained, so we are not able to evaluate the secrecy rate.
As an alternative, we propose to maximize the SNR of the LU, using the zero-forcing beamforming, i.e., we only transmit signal in the null-space of $\check{\bm{h}}_E$.
Assume that the equivalent channel coefficient can be obtained by the LU as in Section \ref{PSASecureTransmission}, then the SNR at the LU is $\mathrm{SNR}_L=\frac{p_B\left|\bm{v}^H\left(\hat{\bm{h}}_L+\tilde{\bm{h}}_L\right)\right|^2}{\sigma_L^2}$. When designing $\bm{v}$, $\tilde{\bm{h}}_L$ is not known by the BS, and hence we use the average SNR as the  objective function. We have
\begin{align}
\label{BeamDesign}
\begin{aligned}
&\mathop{\mathrm{max}}\limits_{\bm{v}\in\mathbb{C}^{ M \times 1}} &&\mathbb{E}_{\tilde{\bm{h}}_L}
\left\{\mathrm{SNR}_L\right\} =\frac{p_B}
{\sigma_L^2}\bm{v}^H\bar{\pmb{H}}_{\mathrm{PJA}}\bm{v},
\\
&\quad\mathrm{s.t.}&& \bm{v}^H\bm{v}=1,\  \check{\bm{h}}_E^H\bm{v}=0.
\end{aligned}
\end{align}
where $\bar{\pmb{H}}_{\mathrm{PJA}}\triangleq\hat{\bm{h}}_L\hat{\bm{h}}_L^H+\tilde{\bm{R}}_{\mathrm{PJA},L}$.
In fact, \eqref{BeamDesign} is equivalent to maximizing $\tilde{\bm{v}}^H\bm{P}^H\bar{\pmb{H}}_{\mathrm{PJA}}
\bm{P}\tilde{\bm{v}}$ with respect to $\tilde{\bm{v}}$, where $\tilde{\bm{v}}\in\mathbb{C}^{\left(M-1\right)\times 1}$ satisfies $\left\|\tilde{\bm{v}}\right\|=1$, and $\bm{P}\in \mathbb{C}^{M\times\left(M-1\right)}$ is a sub-unitary matrix satisfying $\bm{P}^H\check{\bm{h}}_E=\bm{0}$ and $\bm{P}^H\bm{P}=\bm{I}_{M-1}$.
Obviously, we have
$\tilde{\bm{v}}_{\mathrm{opt}}= \bm{u}_{\mathrm{max}}\left\{\bm{P}^H\bar{\pmb{H}}_{\mathrm{PJA}}
\bm{P}\right\}$,
and the optimal solution for \eqref{BeamDesign} is given by $\bm{v}_{\mathrm{opt}}=\bm{P}\tilde{\bm{v}}_{\mathrm{opt}}$.

\section{Numerical Results}
\label{Sectionpart}
In this section, we evaluate the secrecy performance of the proposed uplink channel training and downlink data transmission framework. The BS is equipped with an uniform linear array (ULA) with antenna spacing half of the wavelength.
The following spatially correlated channel model is used in simulations unless specified:
we set $\bm{R}_q=\int_{-\frac{\pi}{2}}^{\frac{\pi}{2}} P_q\left(\theta\right)\left(\bm{a}\left(\theta\right)\right)^H\bm{a}\left(\theta\right) \mathrm{d}\theta$ for $q\in\{L,E\}$, where $\bm{a}\left(\theta\right)\triangleq\left[1,\mathrm{e}^{-\mathrm{j}\pi\sin\left(\theta\right)},\cdots,
\mathrm{e}^{-\mathrm{j}\pi\left(M-1\right)\sin\left(\theta\right)}\right]$ is the steering vector of ULA, $P_q\left(\theta\right)\triangleq \xi_q\mathbb{I}_{\{\theta\in \Theta_q\}}$ for $q\in\{L,E\}$ are the power azimuth spectrum of the LU and Eve, respectively, in which $\xi_q$ is a constant satisfying $\int_{-\frac{\pi}{2}}^{\frac{\pi}{2}} P_q\left(\theta\right)\mathrm{d}\theta=1$, $\Theta_q\triangleq\bigcup_{i=1}^{T_q}
\left[\bar{\theta}_{q,i}-\theta_{q,i}^{(D)}/2,
\bar{\theta}_{q,i}+\theta_{q,i}^{(D)}/2\right]$, $T_L$ ($T_E$) are the numbers of multi-paths from the BS to the LU (Eve), and for $q\in\{L,E\}$, $\bar{\theta}_{q,i}$ and $\theta_{q,i}^{(D)}$ are the center and the angle spread of the $i^{\mathrm{th}}$ path.
Note that when $T_q=1$, this channel model degrades to that in \cite{A.Adhikary2013TIT,L.You2015TWC}.
\subsection{EDR evaluation}
\begin{figure}[t]
    \centering
    \subfigure[$\mathcal{P}_{\mathrm{EDR}}^{(1,2)}$ versus the transmit power of Eve.]{
    \label{EDR:sub1} 
    \includegraphics[width=3 in]{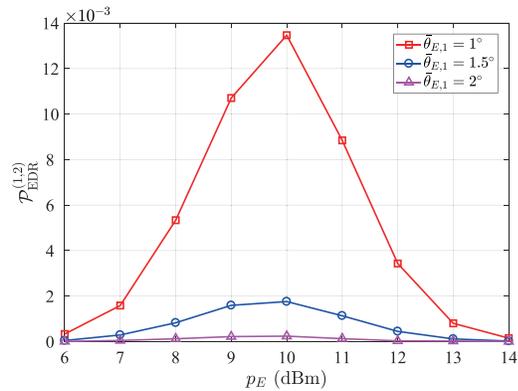}}
    \hspace{.1in}
    \subfigure[$\tilde{\mathcal{P}}_{\mathrm{EDR}}^{(2,2)}$ and $\mathcal{P}_{\mathrm{EDR}}^{(2,2)}$ versus the transmit power of Eve.]{
    \label{EDR:sub2} 
    \includegraphics[width=3 in]{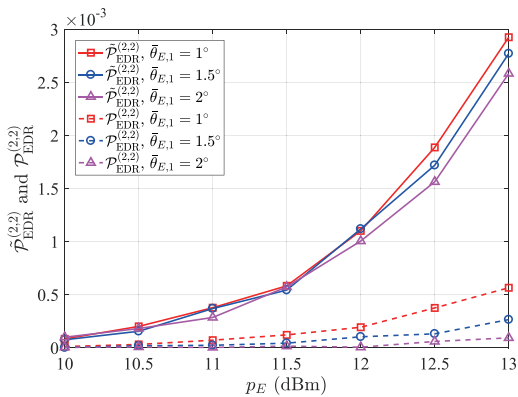}}
    \caption{EDRs under the PSA when there are only two effective channel observations.}
    \label{EDR1}
    \vspace{-5mm}
\end{figure}
\begin{figure}[t]
    \centering
    \subfigure[CCDFs of $d_1^{(C)}$,$d_1^{(N)}$, $\hat{d}_1$, $d_n$, and $\hat{d}_n$.]{
    \label{EDR:sub3} 
    \includegraphics[width=3 in]{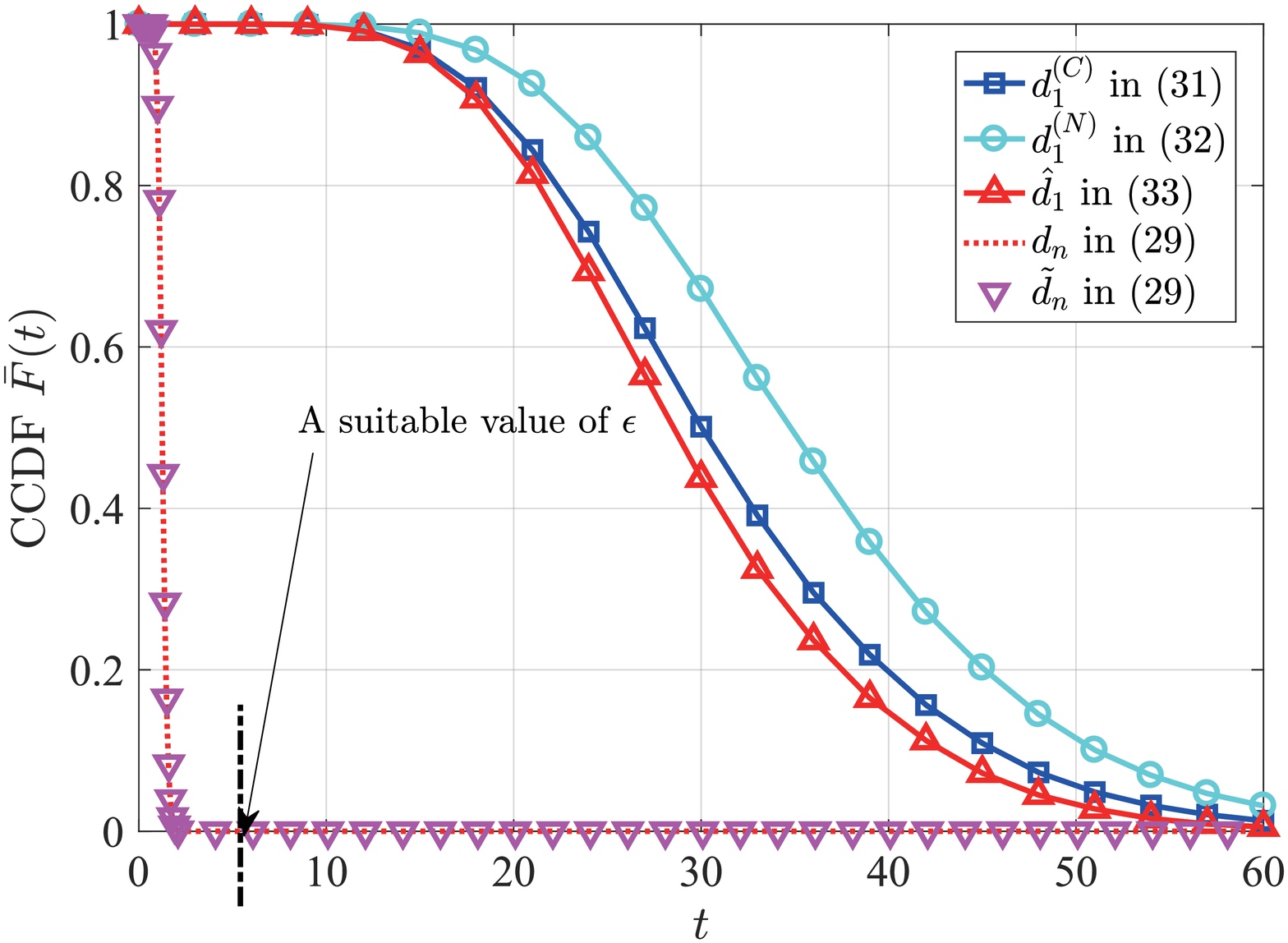}}
    \hspace{.1in}
    \subfigure[EDR versus $K$ under different transmit power of Eve.]{
    \label{EDR:sub4} 
    \includegraphics[width=3 in]{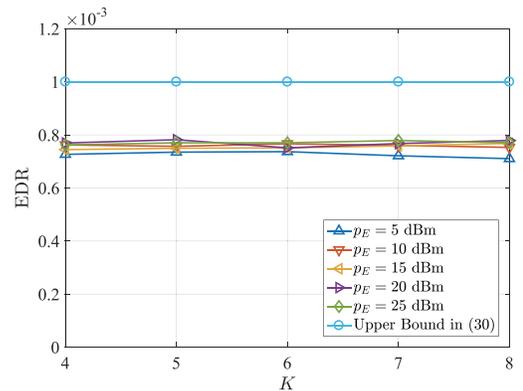}}
  \caption{EDRs under the PSA when there are more than two effective channel observations.}
  \label{EDR2} 
\vspace{-5mm}
\end{figure}
In Fig. \ref{EDR1} and Fig. \ref{EDR2}, we evaluate the EDRs under the PSA through simulation. In Fig. \ref{EDR1}, we simulate the case when there are only two effective channel observations, and in Fig. \ref{EDR2}, we simulate the case when there are more than two effective channel observations. The reason why we simulate the two cases separately is that we use totally different methods to deal with these two cases as shown in Section \ref{PSAChannelEstimation}.

In Fig. \ref{EDR1}, the number of antennas at the BS, the length of the PS, and the transmit power of the LU are set to $M=64$, $\tau=5$, and $p_L=10~\mathrm{dBm}$ , respectively.
We adopt the spatially correlated channel model introduced at the beginning of this section by setting $T_L=T_E=1$, $\bar{\theta}_{L,1}=0^\circ$, and $\theta_{L,1}^{(D)}=\theta_{E,1}^{(D)}=30^\circ$.
In Fig. \ref{EDR:sub1}, we evaluate the EDRs under the condition that Eve only transmits one PS ($K=1$) but fails to hit the PS transmitted by the LU.
As shown in Fig. \ref{EDR:sub1}, for fixed transmit power of Eve,  when the direction of Eve gets closer to that of the LU, the EDR will significantly increase. This is because in this case, the statistical properties, e.g., the covariance matrix, of the legitimate and the illegitimate channels tend to be consistent, which makes it harder to differentiate the legitimate channel from the illegitimate channel. Besides, we show that the value of $\mathcal{P}_{\mathrm{EDR}}^{(1,2)}$
becomes large when $p_E$ approaches to $p_L$ and tends to be small when $p_E$ is quite different from $p_E$. This is because if $p_E$ is much smaller or much larger than $p_L$, the amplitude of $\bm{y}^{(E)}$ in \eqref{L1Psi2y1y2H0H1} will be markedly distinct from that of $\bm{y}^{(L)}$, which makes $\bm{y}^{(L)}$ more distinguishable and thus results in the decreasing EDRs.
In Fig. \ref{EDR:sub2}, we evaluate the EDR under the condition that Eve transmits two PS ($K=2$) and  successfully hits the PS transmitted by the LU. The performance of both the power comparison based method and the GLLR based method are checked. As we can see from Fig. \ref{EDR:sub2}, the GLLR based method generally outperforms the power comparison based method. This is due to the fact that in the power comparison based method, we only utilize partial information of the channel observations, i.e., the power, to determine $\bm{y}^{(L)}$,  while in the GLLR based method, high order statistics of $\bm{y}^{(L)}$ and $\bm{y}^{(E)}$, i.e., the covariance matrices, are utilized to determine $\bm{y}^{(L)}$ and thus enjoys a lower EDR. Besides, we also observe that $\mathcal{P}_{\mathrm{EDR}}^{(2,2)}$ tends to increase with $p_E$. This is because when $p_E$ becomes large, the statistics of $\bm{y}^{(L)}$ will be dominated by that of the term $\mathrm{e}^{\mathrm{j}\omega^{(E)}}\beta_{\mathrm{PSA}}^{(2)}\bm{h}_E$ in \eqref{L2Psi2y1y2H0}, and consequently, the impact of $\bm{h}_L$ on $\bm{y}^{(L)}$  becomes relatively smaller, which means that the statistical differences between $\bm{y}^{(L)}$ and $\bm{y}^{(E)}$ also gets relatively smaller. And therefore, it is harder to distinguish $\bm{y}^{(L)}$ from $\bm{y}^{(E)}$ when $p_E$ is large.

In Fig. \ref{EDR2}, we evaluate the EDRs when there are more than two effective channel observations.
In the simulation, we set $M=32$, and other parameters are set to be the same as those in Fig. \ref{EDR1}, unless specified. Note that the proposed distance based method does not utilize the statistical properties of the legitimate and illegitimate channels to identify $\bm{y}^{(L)}$, and therefore, we set $\bar{\theta}_{E,1}=\bar{\theta}_{L,1}=0^\circ$ for simplicity.
In Fig. \ref{EDR:sub3}, we keep $\beta_{\mathrm{PSA}}^{(K)}=0.5$ regardless of the value of $K$, and we plot the CCDFs of $d_1^{(C)}$, $d_1^{(N)}$, $\hat{d}_1$, $d_n$, and $\tilde{d}_n$ to validate the effectiveness of the approximations in \eqref{dnApproximate} and \eqref{d1LowerBound} and to provide a guideline on designing the decision threshold $\epsilon$.
As shown in Fig. \ref{EDR:sub3}, $\tilde{d}_n$ approximates $d_n$ very well and we can safely use the CCDF of $\tilde{d}_n$, i.e., $\tilde{\mathcal{P}}_F\left(\epsilon\right)$, to approximate $\mathcal{P}_F\left(\epsilon\right)$ as stated in Theorem \ref{DecisionThreshold1}.
To determine a proper value of $\epsilon$, we can choose those values of $\epsilon$ such that $\mathcal{P}\{\tilde{d}_n>\epsilon\}$ is sufficiently small while $\mathcal{P}\{\hat{d}_1<\epsilon\}$ is close to $1$. In this way, the EDR can be controlled to be small.
In Fig. \ref{EDR:sub4}, we set $\tau = 8$, and  we plot the EDRs against the numbers of PSs transmitted by Eve, i.e., $K$, under different power budget of Eve. We choose $\epsilon$ such that $\tilde{\mathcal{P}}_F\left(\epsilon\right)=10^{-3}$.
As we can see from \ref{EDR:sub4}, with the proposed distance-based method, the BS can successfully identify $\bm{y}^{(L)}$ from the effective channel observation set while keeps the EDR to be generally small. We also observe that even if the Eve increases its power or transmits more PSs, the EDR appears to be robust and does not change with the transmission strategies of Eve.

\begin{figure}[t]
\begin{center}
\subfigure[CCDF of $d_n^{(+)}$.]{
    \label{dJA:sub1} 
    \includegraphics[width=3 in]{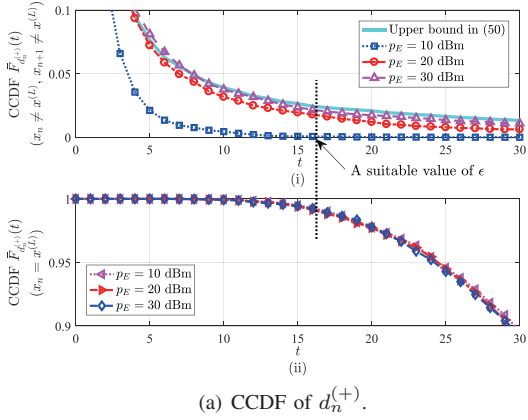}}
    \hspace{.1in}
\subfigure[EDR versus the number of antennas at BS.]{
    \label{dJA:sub2} 
    \includegraphics[width=3 in]{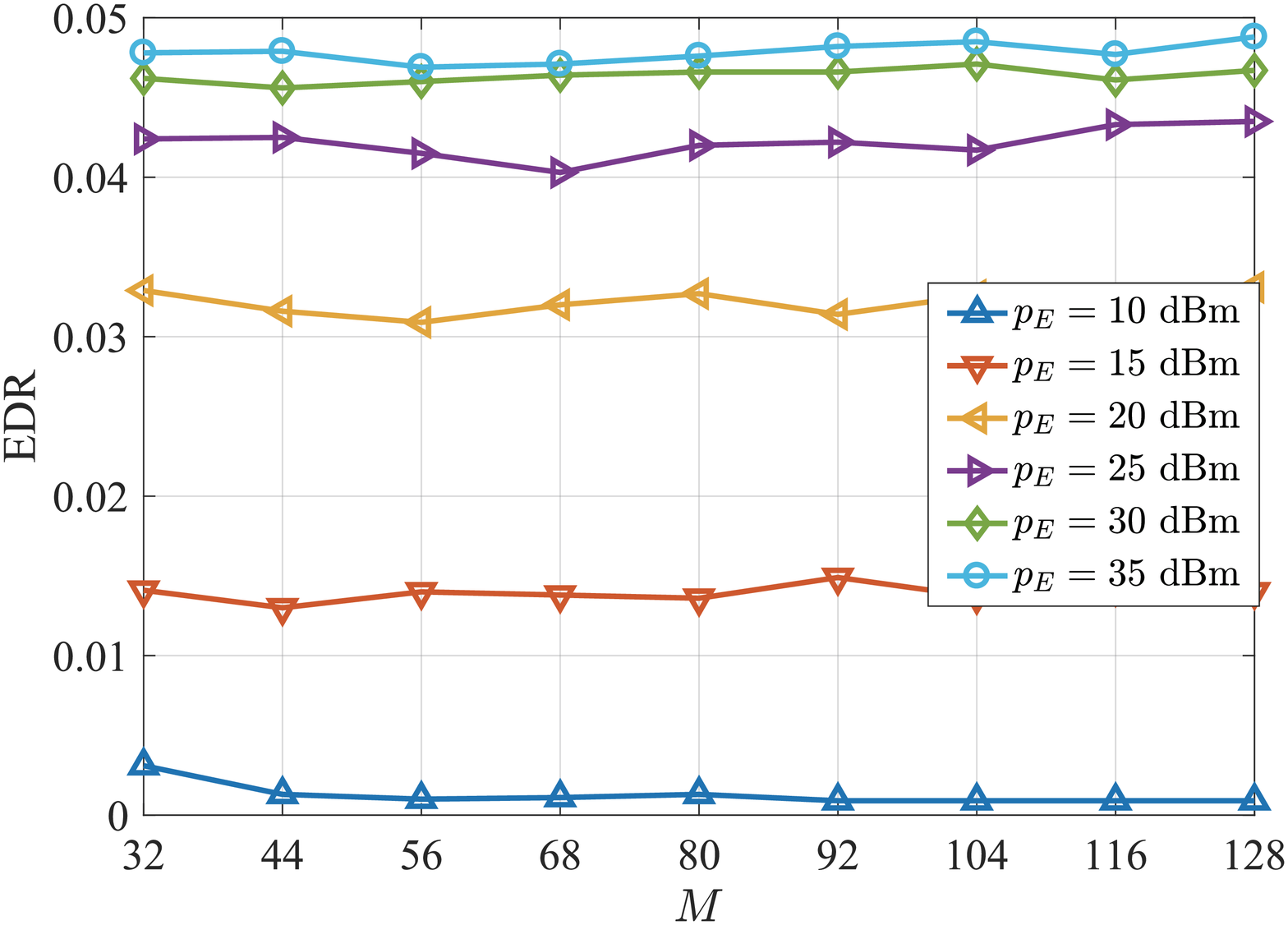}}
\caption{EDRs under the PJA.}\label{Fig:dJA}
\end{center}
\vspace{-5mm}
\end{figure}
In Fig. \ref{Fig:dJA}, the EDRs under PJA are evaluated.
In the simulation, the parameters are set to be the same as those in Fig. \ref{EDR2}, unless specified.
In Fig. \ref{dJA:sub1}, we set $p_L = 15~\mathrm{dBm}$ and $\tau = 5$, and we plot the CCDFs of $d_{n}^{(+)}$ when
$\bm{x}_n\neq\bm{x}^{(L)}$ and $\bm{x}_{n+1}\neq\bm{x}^{(L)}$, and when $\bm{x}_n=\bm{x}^{(L)}$, respectively, to provide a guideline on designing the decision threshold $\epsilon$.
As shown in Fig. \ref{dJA:sub1}, the analytical upper bound accurately approximates the simulation results if $p_E$ is large. In this case, we can directly use $\mathcal{P}_{\mathrm{PJA}}^{(\mathrm{F})}\left(\epsilon\right)$ in \eqref{DJAapproximate} to design $\epsilon$. When $p_E$ is small, $\mathcal{P}_{\mathrm{PJA}}^{(\mathrm{F})}\left(\epsilon\right)$ becomes a strictly upper bound on the simulation results, and in this case, we do not need to restrict a small $\eta$ due to the gap between the upper bound and the exact value. In Fig. \ref{dJA:sub2}, we plot the EDRs versus the numbers of antennas at the BS under different transmit power of Eve. We set $\eta=0.025$ in the simulation. From Fig. \ref{dJA:sub2}, we observe that even though the EDR increases with transmit power of Eve, the EDR still keeps small when the power of the Eve becomes very high (lower than $0.05$ when $p_E=35~\mathrm{dBm}$), which validates the effectiveness of the proposed distance-based method.

We have to emphasize here that ensuring a small EDR is very important for the proposed downlink secure transmission scheme. This is because only when the BS successfully recognizes $\bm{y}^{(L)}$ can it estimates the channel of the LU. If a small EDR is not guaranteed, it is possible that the BS mistakes the illegitimate channel for the legitimate channel, which will undoubtedly result in the private information leakage to Eve.

\subsection{Secrecy performance}
\begin{figure}[t]
\begin{center}
\subfigure[Secrecy rates versus the powers of BS.]{
\label{SecrecyratePSA:sub1} 
\includegraphics[width=3 in]{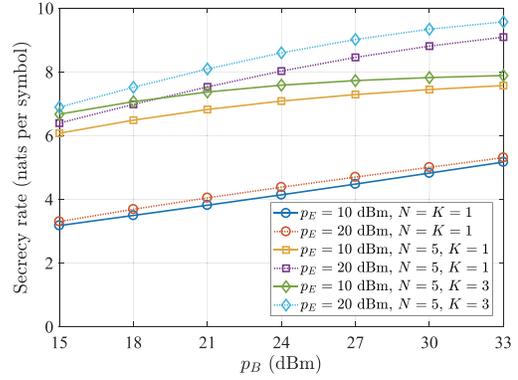}}
\hspace{.1in}
\subfigure[Secrecy rates versus the powers of BS.]{
\label{SecrecyratePSA:sub2} 
\includegraphics[width=3 in]{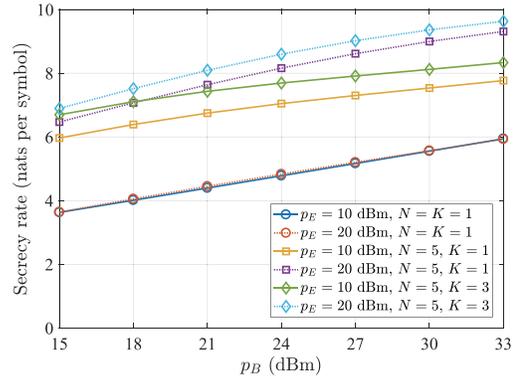}}
\caption{Secrecy performance under the PSA.}\label{Fig:SecrecyratePSA}
\end{center}
\vspace{-3mm}
\end{figure}
We have already shown that the BS can detect $\bm{x}^{(L)}$ with a low EDR in the previous subsection. In this subsection, we evaluate the secrecy performance in terms of secrecy rate under the proposed downlink secure transmission framework.

In Fig. \ref{Fig:SecrecyratePSA}, we plot the secrecy rates versus the downlink transmit powers of the BS when Eve carries out PSA.
In the simulation,  we set $M=48$, $\tau=5$, $p_L=10~\mathrm{dBm}$. The simulation results are obtained by averaging over $1000$ random channel realizations. In each random trial, if the BS makes an error decision on $\bm{y}^{(L)}$, we directly set the secrecy rate as $0$.
In Fig. \ref{SecrecyratePSA:sub1}, we set $T_L=T_E=1$, $\bar{\theta}_{L,1}=0^\circ$, $\bar{\theta}_{E,1}=2^\circ$, and $\theta_{L,1}^{(D)}=\theta_{E,1}^{(D)}=30^\circ$.
In Fig. \ref{SecrecyratePSA:sub2}, we set $T_L=T_E=2$, $\theta_{L,1}^{(D)}=\theta_{E,1}^{(D)}=\theta_{L,2}^{(D)}=\theta_{E,2}^{(D)}=15^\circ$, and $\bar{\theta}_{L,1}=\bar{\theta}_{E,1}=30^\circ$, and in each trial, $\bar{\theta}_{L,2}$ and $\bar{\theta}_{E,2}$ are independently and randomly generated within $\left(-5^\circ,5^\circ\right)$.
In Fig. \ref{Fig:SecrecyratePSA}, $N=1$ corresponds to the conventional training scheme wherein only one PS is allocated to the LU,
and $N=5$ corresponds to the proposed training scheme wherein $5$ PSs are simultaneously allocated to the LU.
As we can see, compared to the conventional training scheme, allocating multiple PSs to the LU can greatly improve the secrecy rate.
This is because when $N=1$, Eve can always hit the LU's PS, which leads to a poor performance of channel estimation. However, when $N>1$, due to the fact that Eve does not know which PS the LU will transmit, it is possible for the BS to obtain uncontaminated channel observations of both the LU's and Eve's channels, which significantly improves the accuracy of the channel estimations, and thus the secrecy rate gets increased. Besides, we observe that the secrecy rate is improved when the transmit power of Eve is increased from $10~\mathrm{dBm}$ to $20~\mathrm{dBm}$. Note that this phenomenon is totally different from that shown in \cite{X.Zhou2012TWC}. In \cite{X.Zhou2012TWC}, the authors showed that without protection on the channel training procedure, Eve can always improve its wiretapping performance by increasing its power. However, with the proposed channel training scheme, if Eve increases its transmit power during the channel training procedure, the BS will obtain more accurate estimation of the illegitimate channel, which undoubtedly degrades the wiretapping capability of Eve.

\begin{figure}[t]
\begin{center}
\subfigure[Achieve rates versus the powers of Eve.]{
    \label{SecrecyrateJA:sub1} 
    \includegraphics[width=3 in]{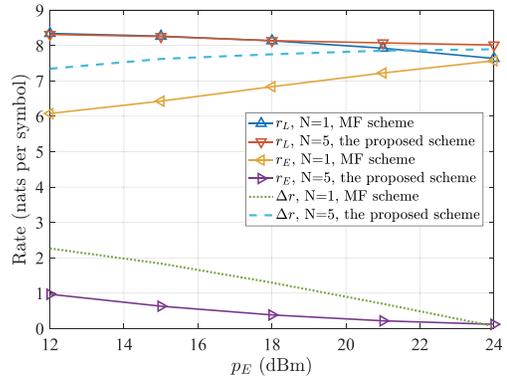}
    }
    \hspace{.1in}
\subfigure[Achieve rates versus the powers of Eve.]{
    \label{SecrecyrateJA:sub2} 
    \includegraphics[width=3 in]{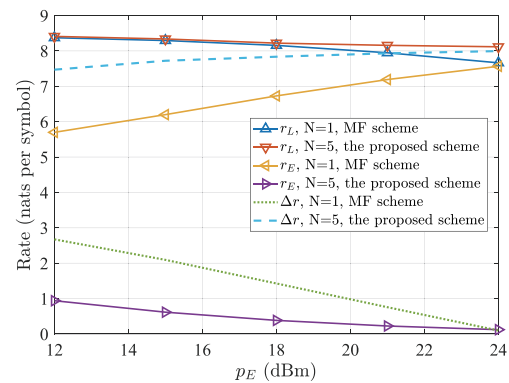}}
\caption{Secrecy performance under PJA.}\label{Fig:SecrecyrateJA}
\end{center}
\vspace{-5mm}
\end{figure}

In Fig. \ref{Fig:SecrecyrateJA}, we check the performance of the proposed secure downlink transmission framework under PJA. In the simulation, we set $M=48$, $\tau = 5$, $p_L=15$ (dBm), $p_B=20$ (dBm), and $\eta=0.025$. Besides, we consider an extreme terrible case for the LU where $\bm{R}_L=\bm{R}_E$.
In Fig. \ref{SecrecyrateJA:sub1}, the spatially correlated channel model is adopted by setting $T_L=T_E=1$, $\bar{\theta}_{L,1} = \bar{\theta}_{E,1}=0^\circ$, and $\theta_{L,1}^{(D)} = \theta_{E,1}^{(D)}=30^\circ$. In Fig. \ref{SecrecyrateJA:sub2}, we set $\bm{R}_L=\bm{R}_E=\bm{I}_M$.
The achievable rates of the LU and Eve, denoted by $r_L\triangleq\ln(1+\mathrm{SNR_L})$ and $r_E\triangleq\ln(1+\mathrm{SNR_E})$, respectively, and the differences of the achievable rates, denoted by  $\Delta_r\triangleq r_L-r_E$, are used as the performance metrics.
We also provide the performance of the conventional channel training ($N=1$) and matching filter (MF) beamforming data transmission scheme for comparison.
The simulation results are obtained by averaging across $1000$ random channel realizations.
As we can see in Fig. \ref{Fig:SecrecyrateJA}, in the conventional channel training and MF beamforming transmission scheme, $r_E$ is generally very high, and even approaches to $r_L$ when $p_E=24~\mathrm{dBm}$.
This means that the data transmission suffers from serious signal leakage. The main reason accounts for this phenomenon is that under the condition $\bm{R}_L=\bm{R}_E$, the BS can hardly distinguish the legitimate channel from the wiretap channel, and therefore, there is no way for the BS to restrain the signal leakage.
However, in the proposed channel training and data transmission framework,
$r_E$ is always much lower than $r_L$, and even approaches to $0$ when $p_E=24~\mathrm{dBm}$.
This is because with the proposed channel training scheme, the BS is able to simultaneously estimate the legitimate channel and the channel direction of the wiretap channel, based on which zero-forcing beamforming can be utilized to greatly reduce the signal power received by Eve. We also observe  that $r_L$ slightly decreases with the increase of $p_E$. This is because when $p_E$ gets large, the EDR also increase as shown in Fig. \ref{dJA:sub2}. This means that the probability that the BS mistakes the illegitimate channel as the legitimate channel increases, and therefore $r_L$ decreases accordingly.

\section{Conclusion}
In this paper, a framework, including the RCT scheme and the SB vector design, was proposed to combat with the PSA and PJA during the channel training phase in a TDD system. We hid LU's PS by letting the LU randomly select a PS to transmit.
The BS can recognize LU's PS by utilizing the different channel statistics between Eve and the LU.
It was shown that though the BS does not know which PS will be transmitted by LU in advance, it can correctly recognize it with very high probability, even the channel statistics of Eve is very similar to that of the LU. We showed that the secrecy rate is remarkably increased when the proposed framework is utilized.

At last, we have to point out that in this paper, we only investigated the single-cell systems.
When it comes to multi-cell systems, the PSs will be reused in multiple cells, and the pilot contamination from the other cells will makes the problem much more difficult and complicated. To combat PSA and PJA in multi-cell situations will be one of our future works.

\appendix

\subsection{The Proof of Theorem \ref{Theorem:T22}}
\label{Appendix:ProofT22}
Define a unitary matrix $\bm{\Lambda}\left(x\right)\triangleq\mathrm{diag}\left(\bm{I}_M,\mathrm{e}^{\mathrm{j}x}\bm{I}_M\right)
$.
Because $\bm{h}_E$ is circular and symmetrical, we have
$\tilde{\mathcal{P}}_{\mathrm{EDR}}^{(2,2)}
=\mathcal{P}\left\{\left\|\tilde{\bm{g}}_1\right\|^2 -
\left\|\tilde{\bm{g}}_2\right\|^2<0\right\}$, where $\tilde{\bm{g}}_1\triangleq\bm{h}_L + \beta_{\mathrm{PSA}}^{(2)}\bm{h}_E + \bm{z}^{(L)}$, $\tilde{\bm{g}}_2\triangleq\mathrm{e}^{\mathrm{j}\tilde{\omega}}\beta_{\mathrm{PSA}}^{(2)}\bm{h}_E+\bm{z}^{(E)}$, and $\tilde{\omega}\triangleq\omega^{(E)}-\omega^{(L)}$.
Using $\bm{\Lambda}\left(x\right)$, we have
$\tilde{\mathcal{P}}_{\mathrm{EDR}}^{(2,2)}=\mathcal{P}\left\{\tilde{\bm{g}}^H\bm{\Lambda}\left(\pi\right)
\tilde{\bm{g}}<0\right\}$,
where $\tilde{\bm{g}}_\triangleq\left[\tilde{\bm{g}}_1^H,\tilde{\bm{g}}_2^H\right]^H
\sim\mathbb{CN}\left(\bm{0},\bm{R}_g\left(\tilde{\omega}\right)\right)$ with
\begin{align}
\bm{R}_g\left(\tilde{\omega}\right)&\triangleq\mathbb{E}\left(\tilde{\bm{g}}\tilde{\bm{g}}^H\right)=
\begin{bmatrix}
\bm{R}_{L,E,z}^{(2)}& \mathrm{e}^{-\mathrm{j}\tilde{\omega}}\bm{R}_E^{(2)}\\
\mathrm{e}^{\mathrm{j}\tilde{\omega}}\bm{R}_E^{(2)}&
\bm{R}_{E,z}^{(2)}
\end{bmatrix}\nonumber \\
&=\bm{\Lambda}\left(\tilde{\omega}\right)
\bm{R}_g\left(0\right)
\bm{\Lambda}\left(-\tilde{\omega}\right).
\end{align}
Therefore, we further have
\begin{align}
\tilde{\mathcal{P}}_{\mathrm{EDR}}^{(2,2)}& =
\mathcal{P}\Big\{
\tilde{\bm{g}}^H
\left(\bm{R}_g\left(\tilde{\omega}\right)\right)^{-\frac{1}{2}}
\left(\bm{R}_g\left(\tilde{\omega}\right)\right)^{\frac{1}{2}}
\bm{\Lambda}\left(\pi\right)
\nonumber \\
&\quad\quad\quad \times\left(\bm{R}_g\left(\tilde{\omega}\right)\right)^{\frac{1}{2}}
\left(\bm{R}_g\left(\tilde{\omega}\right)\right)^{-\frac{1}{2}}
\tilde{\bm{g}}
<0\Big\}\nonumber\\
& = \mathcal{P}\Big\{
\tilde{\bm{g}}^H\bm{\Lambda}\left(\tilde{\omega}\right)
\left(\bm{R}_g\left(0\right)\right)^{-\frac{1}{2}}
\left(\bm{R}_g\left(0\right)\right)^{\frac{1}{2}}
\bm{\Lambda}\left(\pi\right)
\nonumber \\
&\quad\quad\quad \times\left(\bm{R}_g\left(0\right)\right)^{\frac{1}{2}}
\left(\bm{R}_g\left(0\right)\right)^{-\frac{1}{2}}
\bm{\Lambda}\left(-\tilde{\omega}\right)\tilde{\bm{g}}
<0\Big\}\nonumber\\
&=\mathcal{P}\left\{
\bm{\alpha}^H
\bm{W}
\bm{\alpha}
<0\right\},
\label{S4}
\end{align}
where $\bm{\alpha}\triangleq\left(\bm{R}_g\left(0\right)\right)^{-\frac{1}{2}}
\bm{\Lambda}\left(-\tilde{\omega}\right)\tilde{\bm{g}}$ and
$\bm{W}\triangleq \left(\bm{R}_g\left(0\right)\right)^{\frac{1}{2}}
\bm{\Lambda}\left(\pi\right)
\left(\bm{R}_g\left(0\right)\right)^{\frac{1}{2}}$.
Note that $\bm{W}$ does not depend on $\omega$.
As for $\bm{\alpha}$, it can be easily checked that
$\mathbb{E}\left\{\bm{\alpha}\bm{\alpha}^H\right\}
=\pmb{I}_{2M}$.

\subsection{The proof Lemma \ref{Approximation1}}
\label{Lemma1Lemma2Proof}
We first show that $\mathbb{E}_{X,Y}\left\{\ln\left(\frac{1+X}{1+Y}\right)\right\}$ and $\ln\left(\frac{1+\mathbb{E}_{X}\left\{X\right\}}{1+\mathbb{E}_{Y}\left\{Y\right\}}\right)$
have a common lower bound as follows,
\begin{align}
&\mathbb{E}_{X,Y}\left\{\ln\left(\frac{1+X}{1+Y}\right)\right\}\nonumber\\
&\overset{(a)}{\geq}
\ln\left(1+\frac{1}{\mathbb{E}_X\left\{1/X\right\}}\right)
-\ln\left(1+\mathbb{E}_{Y}\left\{Y\right\}\right)\nonumber\\
&\overset{(b)}{\leq}
\ln\left(\frac{1+\mathbb{E}_X\left\{X\right\}}{1+\mathbb{E}_{Y}\left\{Y\right\}}\right),
\label{Bound1}
\end{align}
where step $(a)$ is obtained by applying Jensen's inequality to convex functions $\ln\left(1+\frac{1}{x}\right)$ and $\ln\left(\frac{1}{1+x}\right)$,
and step $(b)$ is because $\mathbb{E}_{X}\left\{\frac{1}{X}\right\}\geq \frac{1}{\mathbb{E}_{X}\left\{X\right\}}$.
Now we show that $\mathbb{E}_{X,Y}\left\{\ln\left(\frac{1+X}{1+Y}\right)\right\}$ and $\ln\left(\frac{1+\mathbb{E}_{X}\left\{X\right\}}{1+\mathbb{E}_{Y}\left\{Y\right\}}\right)$
have a common upper bound as follows,
\begin{align}
&\mathbb{E}_{X,Y}\left\{\ln\left(\frac{1+X}{1+Y}\right)\right\}\nonumber\\
&
\overset{(a)}{\leq}
\ln\left( 1+\mathbb{E}_{X}\left\{X\right\}\right) - \ln\left(1+\frac{1}{\mathbb{E}_Y\left\{1/Y\right\}}\right)\nonumber\\
&\overset{(b)}{\geq}
\ln\left(\frac{1+\mathbb{E}_X\left\{X\right\}}{1+\mathbb{E}_{Y}\left\{Y\right\}}\right),\label{Bound2}
\end{align}
where step $(a)$ is due to the fact that $\ln\left(1+x\right)$ and $-\ln\left(1+\frac{1}{x}\right)$ are concave functions, and step $(b)$ is because
$\mathbb{E}_{Y}\left\{\frac{1}{Y}\right\}\geq \frac{1}{\mathbb{E}_{Y}\left\{Y\right\}}$.
As we can see, $\mathbb{E}_{X,Y}\left\{\ln\left(\frac{1+X}{1+Y}\right)\right\}$ and $\ln\left(\frac{1+\mathbb{E}_{X}\left\{X\right\}}{1+\mathbb{E}_{Y}\left\{Y\right\}}\right)$ has the same upper and lower bound, and thus we can use
$\ln\left(\frac{1+\mathbb{E}_{X}\left\{X\right\}}{1+\mathbb{E}_{Y}\left\{Y\right\}}\right)$ to approximate
$\mathbb{E}_{X,Y}\left\{\ln\left(\frac{1+X}{1+Y}\right)\right\}$.

\subsection{Proof of Theorem \ref{JammingAttackUpperbound}}
\label{Appendix:ProofUpperBound}
The derivation of Theorem \ref{JammingAttackUpperbound} is given as follows,
\begin{align}
&\mathcal{P}\left\{d_{\mathrm{PJA},n}^{(+)}>\epsilon\right\}
\nonumber\\
&=\mathcal{P}\Big\{\mathop{\mathrm{min}}\limits_a ~
\|\mu_n\beta_{\mathrm{PJA}}\bm{h}_E + \bm{z}_n\nonumber\\
&
\quad- a\left(\mu_{n+1}\beta_{\mathrm{PJA}}\bm{h}_E + \bm{z}_{n+1}\right)\|^2>\epsilon\Big\}
\nonumber\\
&\overset{(a)}{\leq}
\mathcal{P}\left\{\left\|\bm{z}_n - \mu_n\bm{z}_{n+1}/\mu_{n+1}\right\|^2>\epsilon\right\}\nonumber\\
&\overset{(b)}{=}\tau^2\int_0^{+\infty}\int_0^{+\infty}
\mathrm{e}^{-\tau\left(x+y\right)}\frac{\Gamma\left(M,\frac{\epsilon x}{\sigma_z^2\left(x+y\right)}\right)}{\Gamma\left(M\right)}\mathrm{d}x\mathrm{d}y\nonumber \\
&\overset{(c)}{=}\tau^2\int_0^{+\infty}\int_v^{+\infty}
\mathrm{e}^{-\tau u}\frac{\Gamma\left(M,\frac{\epsilon v}{\sigma_z^2u}\right)}{\Gamma\left(M\right)}\mathrm{d}u\mathrm{d}v\nonumber\\
&\overset{(d)}{=}\tau^2\int_0^{+\infty}\int_0^{u}
\mathrm{e}^{-\tau u}\mathrm{e}^{-\frac{\epsilon }{\sigma_z^2u}v}\sum_{m=0}^{M-1}
\frac{\left(\frac{\epsilon }{\sigma_z^2u}\right)^mv^m}{m!}\mathrm{d}v\mathrm{d}u\nonumber \\
&\overset{(e)}{=}\tau^2\int_0^{+\infty}\mathrm{e}^{-\tau u}\sum_{m=0}^{M-1}\left\{
\frac{\sigma_z^2u}{\epsilon}-\sum_{k=0}^m\frac{u^k\left(\frac{\epsilon }{\sigma_z^2u}\right)^{k-1}}
{\mathrm{e}^{\frac{\epsilon}{\sigma_z^2}}k!}
\right\}\mathrm{d}u\nonumber\\
&=\frac{\sigma_z^2M}{\epsilon}-
\sum_{m=0}^{M-1}\sum_{k=0}^m\frac{\left(\frac{\epsilon }{\sigma_z^2}\right)^{k-1}}{\mathrm{e}^{\frac{\epsilon}{\sigma_z^2}}k!}.
\label{FinalUpper}
\end{align}
In \eqref{FinalUpper},
step $(a)$ is obtained by setting $a=\frac{\mu_n}{\mu_{n+1}}$,
step $(b)$ is because $\left\|\bm{z}_n - \mu_n\bm{z}_{n+1}/\mu_{n+1}\right\|^2
\sim\mathbb{G}\left(M,\frac{\left|\mu_{n+1}\right|^2+\left|\mu_n\right|^2}{\left|\mu_{n+1}\right|^2}\right)$ when $\mu_n$ and $\mu_{n+1}$ are fixed, and $\left|\mu_{n}\right|^2\sim\mathbb{E}\left(\frac{1}{\tau}\right)$ and $\left|\mu_{n+1}\right|^2\sim\mathbb{E}\left(\frac{1}{\tau}\right)$,
step $(c)$ is obtained by $x+y\rightarrow u$ and $x\rightarrow v$,
step $(d)$ is obtained by using \cite[Eq. (3.3512)]{I.S.Gradshteyn2007Table}, and
step $(e)$ is obtained by using \cite[Eq. (3.3511)]{I.S.Gradshteyn2007Table}.
\subsection{CCDF of quadratic formula of Gaussian random vector}
\label{QuadraticGaussian}
We provide an expression for $\mathcal{P}\left\{\bm{x}^H\bm{\Omega}\bm{x}> t\right\}$ in this appendix, where $t\geq0$, $\bm{x}\sim\mathbb{CN}\left(\bm{0},\bm{I}_L\right)$, and $\bm{\Omega}\in\mathbb{C}^{L\times L}$ is an indefinite Hermitian matrix.
Denote $s^{(1,1)}<s^{(1,2)}<\cdots<s^{(1,n_1)}$ as the $n_1$ distinct positive eigenvalues of $\bm{\Omega}$ with algebraic multiplicities given by $m_{1,1},m_{1,2},\cdots,m_{1,n_1}$, respectively.
Denote $s^{(2,n_2)}<s^{(2,n_2-1)}<\cdots<s^{(2,1)}$ as the $n_2$ distinct negative eigenvalues of $\bm{\Omega}$ with algebraic multiplicities given by $m_{2,n_2},m_{2,n_2-1},\cdots,m_{2,1}$, respectively.
We have
\begin{align}
\mathcal{P}\left\{\bm{x}^H\bm{\Omega}\bm{x} \geq t\right\} &=
\mathcal{P}\left\{\bm{x}^H\bm{U}\bm{D}\bm{U}^H\bm{x} \geq t\right\} \nonumber\\
&=\mathcal{P}\left\{\bm{\mu}^H\bm{D}\bm{\mu} \geq t\right\} \nonumber\\
&=\mathcal{P}\left\{ Q_1- Q_2\geq t\right\}\nonumber\\
&=\int_0^{+\infty}\int_{y+t}^{+\infty}f_{Q_1}\left(x\right)f_{Q_2}\left(y\right)
\mathrm{d}x\mathrm{d}y,\nonumber
\end{align}
where $\bm{U}\bm{D}\bm{U}^H=\bm{\Omega}$ is the eigenvalue decomposition, and
\begin{align}
\bm{\mu}&\triangleq\bm{U}^H\bm{x}\sim\mathbb{CN}\left(\bm{0},\bm{I}_L\right),\nonumber\\
Q_1 &\triangleq \sum_{i=1}^{n_1}G_{1,i},\quad Q_2 \triangleq \sum_{j=1}^{n_2}G_{2,j},\nonumber\\
G_{1,i}&\sim \mathbb{G}_{1,i}\left(m_{1,i},s^{(1,i)}\right), i=1,2,\cdots,n_1,\nonumber\\
G_{2,j}&\sim \mathbb{G}_{2,j}\left(m_{2,j},\left|s^{(2,i)}\right|\right), j=1,2,\cdots,n_2.\nonumber
\end{align}
According to \cite{P.G.Moschopoulos1985}, the PDF of $Q_q$, for $q \in\{1,2\}$, can be written as
\begin{align}
f_{Q_q}\left(x\right)=\mathbb{I}_{\{x>0\}}\sum_{k=0}^{\infty}
\frac{\bar{\delta}_{q,k} x^{\rho_q+k-1}}{\left|s^{(q,1)}\right|^{\rho_q+k}\Gamma\left(\rho_q+k\right)}
\mathrm{e}^{-\frac{x}{\left|s^{(q,1)}\right|}},\nonumber
\end{align}
where for $q\in\{1,2\}$,
$\bar{\delta}_{q,k}\triangleq C_q\delta_{q,k}$,
$C_q\triangleq\prod_{i=1}^{n_q}\left(\frac{s^{(q,1)}}{s^{(q,i)}}\right)^{m_{q,i}}$,
$\delta_{q,k+1} \triangleq \frac{1}{k+1} \sum_{i=1}^{k+1}i\gamma_{q,i} \delta_{q,k+1-i}$, $\delta_{q,0} \triangleq 1$,
$\gamma_{q,k} \triangleq \sum_{i=1}^{n_q}\frac{m_{q,i}}{k}\left(1 - \frac{s^{(q,1)}}{s^{(q,i)}}\right)^k$,
and $\rho_q\triangleq\sum_{i=1}^{n_q} m_{q,i}$.
Based on $f_{Q_q}\left(x\right)$, we have
\begin{align}
&\int_0^{+\infty}\int_{y+t}^{+\infty}f_{Q_1}\left(x\right)f_{Q_2}\left(y\right)
\mathrm{d}x\mathrm{d}y \nonumber\\
&=\int_0^{+\infty}
\sum_{k_2=0}^{\infty}
\frac{\bar{\delta}_{2,k_2}y^{\rho_2+k_2-1}\mathrm{e}^{-\frac{y}{\left|s^{(2,1)}\right|}}}
{\left|s^{(2,1)}\right|^{\rho_2+k_2}\Gamma\left(\rho_2+k_2\right)}\nonumber \\
&\quad \times
\int_{y+t}^{+\infty}
\sum_{k_1=0}^{\infty}
\frac{\bar{\delta}_{1,k_1}x^{\rho_1+k_1-1}\mathrm{e}^{-\frac{x}{\left|s^{(1,1)}\right|}}}
{\left|s^{(1,1)}\right|^{\rho_1+k_1}\Gamma\left(\rho_1+k_1\right)}
\mathrm{d}x\mathrm{d}y\nonumber\\
&=\sum_{k_1=0}^{\infty}\sum_{k_2=0}^{\infty}
\bar{\delta}_{1,k_1}
\bar{\delta}_{1,k_2}\Delta_{k_1,k_2}
\left(t\right).\label{Q1minusQ2CDFS3}
\end{align}
where due to the fact that $\rho_1$ and $\rho_2$ are integers, $\Delta_{k_1,k_2}
\left(t\right)$ can be written as \eqref{FurtherSimplifiedDeltat} at the top of the next page.
\begin{figure*}
\begin{align}
\Delta_{k_1,k_2}
\left(t\right)&= 
\int_0^{+\infty}
\frac{y^{\rho_2+k_2-1}\mathrm{e}^{-y}\Gamma\left(\rho_1+k_1,\left|\frac{s^{(2,1)}}{s^{(1,1)}}\right|\left(y+\frac{t}{\left|s^{(2,1)}\right|}\right)\right)}{\Gamma\left(\rho_2+k_2\right)\Gamma\left(\rho_1+k_1\right)}
\mathrm{d}y \nonumber\\
& =\int_0^{+\infty}
\frac{y^{\rho_2+k_2-1}\mathrm{e}^{-y}}
{\Gamma\left(\rho_2+k_2\right)}
\mathrm{e}^{-\left|\frac{s^{(2,1)}}{s^{(1,1)}}\right|\left(y+\frac{t}{\left|s^{(2,1)}\right|}\right)}
\sum_{i=0}^{\rho_1+k_1-1}\frac{1}{i!}\left|\frac{s^{(2,1)}}{s^{(1,1)}}\right|^i
\left(y+\frac{t}{\left|s^{(2,1)}\right|}\right)^i
\mathrm{d}y\nonumber\\
& =\int_0^{+\infty}
\frac{y^{\rho_2+k_2-1}\mathrm{e}^{-y}}
{\Gamma\left(\rho_2+k_2\right)}
\mathrm{e}^{-\left|\frac{s^{(2,1)}}{s^{(1,1)}}\right|\left(y+\frac{t}{\left|s^{(2,1)}\right|}\right)}
\sum_{i=0}^{\rho_1+k_1-1}\frac{1}{i!}\left|\frac{s^{(2,1)}}{s^{(1,1)}}\right|^i
\sum_{l=0}^{i}\binom{l}{i} y^l
\left(\frac{t}{\left|s^{(2,1)}\right|}\right)^{i-l}
\mathrm{d}y\nonumber\\
& =
\sum_{i=0}^{\rho_1+k_1-1}\frac{1}{i!}\left|\frac{s^{(2,1)}}{s^{(1,1)}}\right|^i
\sum_{l=0}^{i}\binom{l}{i}
\left(\frac{t}{\left|s^{(2,1)}\right|}\right)^{i-l}
\mathrm{e}^{-\frac{t}{\left|s^{(1,1)}\right|}}
\int_0^{+\infty}
\frac{y^{\rho_2+k_2+l-1}\mathrm{e}^{-\left(1+\left|\frac{s^{(2,1)}}{s^{(1,1)}}\right|\right)y}}
{\Gamma\left(\rho_2+k_2\right)}
\mathrm{d}y \nonumber\\
& =
\sum_{i=0}^{\rho_1+k_1-1}\frac{\mathrm{e}^{-\frac{t}{\left|s^{(1,1)}\right|}}}{i!}
\left|\frac{s^{(2,1)}}{s^{(1,1)}}\right|^i
\sum_{l=0}^{i}\binom{l}{i}
\left(\frac{t}{\left|s^{(2,1)}\right|}\right)^{i-l}
\frac{\Gamma\left(\rho_2+k_2+l\right)}
{\Gamma\left(\rho_2+k_2\right)}
\left(\frac{\left|s^{(1,1)}\right|}{\left|s^{(1,1)}\right|+\left|s^{(2,1)}\right|}\right)^{\rho_2+k_2+l}.
\label{FurtherSimplifiedDeltat}
\end{align}
\hrulefill
\end{figure*}
Inserting $\Delta_{k_1,k_2}\left(t\right)$ into \eqref{Q1minusQ2CDFS3}, we obtain
\begin{align}
&\mathcal{P}\left\{\pmb{x}^H\pmb{\Omega}\pmb{x}> t\right\}=\sum_{k_2=0}^{\infty}\sum_{k_1=0}^{\infty}
\bar{\delta}_{1,k_1}\bar{\delta}_{1,k_2}
\sum_{i=0}^{\rho_1+k_1-1}\frac{\mathrm{e}^{-\frac{t}{\left|s^{(1,1)}\right|}}}{i!}
\nonumber\\
&\quad\times
\left|\frac{s^{(2,1)}}{s^{(1,1)}}\right|^i\sum_{l=0}^{i}\binom{l}{i}
\left(\frac{t}{\left|s^{(2,1)}\right|}\right)^{i-l}
\frac{\Gamma\left(\rho_2+k_2+l\right)}
{\Gamma\left(\rho_2+k_2\right)}\nonumber\\
&\quad\times
\left(\frac{\left|s^{(1,1)}\right|}{\left|s^{(1,1)}\right|+\left|s^{(2,1)}\right|}\right)^{\rho_2+k_2+l}.
\end{align}
For the special case of $t=0$, we have
\begin{align}
\mathcal{P}\left\{\pmb{x}^H\pmb{\Omega}\pmb{x}> t\right\}&=
\sum_{k_2=0}^{\infty}\sum_{k_1=0}^{\infty}
\bar{\delta}_{1,k_1}\bar{\delta}_{1,k_2}
\sum_{i=0}^{\rho_1+k_1-1}
\frac{1}{i!}\nonumber\\
&\quad \times
\frac{\Gamma\left(\rho_2+k_2+i\right)}
{\Gamma\left(\rho_2+k_2\right)}
\left|\frac{s^{(2,1)}}{s^{(1,1)}}\right|^i\nonumber\\
&\quad \times
\left(\frac{\left|s^{(1,1)}\right|}{\left|s^{(1,1)}\right|+\left|s^{(2,1)}\right|}\right)^{\rho_2+k_2+i}.
\end{align}

\end{document}